\documentclass[aps,prb,twocolumn,groupedaddress,floats,showpacs]{revtex4}
\usepackage{latexsym} 
\usepackage{dcolumn}
\usepackage[dvips]{graphicx} 
\usepackage{amssymb}
\usepackage{graphics}
\usepackage{amsmath}
\usepackage{epsf}

\newcommand{\vecmu}{\mbox{\boldmath $\mu$} }
\newcommand{\vectau}{\mbox{\boldmath $\tau$} }

\newcommand{\vecchi}{\mbox{\boldmath $\chi$} }
\newcommand{\up}{\uparrow}
\newcommand{\down}{\downarrow}
\newcommand{\lsf}{l_{\rm sf}}

\newcommand{\vk}{\vec k}
\newcommand{\va}{\vec a}
\newcommand{\ve}{\vec e}
\newcommand{\vj}{\vec j}
\newcommand{\vm}{\vec m}
\newcommand{\vmd}{\dot{\vm}}

\newcommand{\vq}{\vec q}
\newcommand{\vqp}{{\vec q^{\scriptstyle{\prime}}}}
\newcommand{\vqpp}{{\vec q{\scriptscriptstyle{'}}}}

\renewcommand{\vr}{\vec r}

\newcommand{\vmu}{\mbox{\boldmath $\mu$}}
\newcommand{\vtau}{\mbox{\boldmath $\tau$}}

\newcommand{\Gq}{G_{\rm s}(\vq)}
\newcommand{\Gk}{G_{\rm s}(\vk)}
\newcommand{\Go}{G_{\rm s}(0)}
\newcommand{\Gqp}{G_{\rm s}(\vqpp)}
\newcommand{\Gqc}{G_{\rm s}(\vq_{\rm c})}
\renewcommand{\vec}[1]{\mathbf{#1}}

\begin{document}
\author{Shaffique Adam, Mikhail L.\ Polianski,\footnote{Present
 address: D\'epartement de Physique Th\'eorique, Universit\'e de
Gen\`eve, CH-1211 Gen\`eve 4, Switzerland}
 and Piet W.\ Brouwer}
\title{Current induced transverse spin-wave instability in thin ferromagnets:
beyond linear stability analysis}

\affiliation{Laboratory of Atomic and Solid State Physics,
Cornell University, Ithaca, NY 14853-2501}
\date{\today}
\begin{abstract}
A sufficiently large unpolarized current can cause a spin-wave
instability in thin nanomagnets with asymmetric contacts. The
dynamics beyond the instability is understood in the 
perturbative regime of small spin-wave amplitudes, as well as
by numerically solving a discretized model. In the absence of
an applied magnetic field, our numerical simulations reveal
a hierarchy of instabilities, leading to chaotic magnetization
dynamics for the largest current densities we consider.  
\end{abstract}
\maketitle


\section{Introduction}
\label{sec:1}

Almost a decade ago Slonczewski\cite{kn:slonczewski1996} and Berger
\cite{kn:berger1996} predicted that when a
spin-polarized current is passed through a ferromagnet it transfers the
transverse component of its spin angular momentum to the ferromagnet.
The experimental verification of the theoretical predictions
followed within a few 
years.\cite{kn:tsoi1998,kn:sun1999,kn:wegrowe1999,kn:myers1999,kn:katine2000}
Since then, the so-called `spin-transfer effect' has been observed
in a large number of different experiments.

In most experiments, the spin-transfer torque is studied in a
ferromagnet--normal-metal--ferromagnet
tri-layer structure where a thick ferromagnet first
polarizes the current which then exerts a spin-transfer torque on a
second thinner ferromagnet. At sufficiently large applied current
densities, 
the spin-transfer torque then may alter the magnetization
direction of the thin magnet. The observation of hysteretic magnetic
switching for one current direction only was seen as a hallmark of the
spin-torque effect,\cite{kn:myers1999}
and excluded an explanation of the experiments in
terms of the Faraday field associated with the applied current. 
(Note that for small system sizes, the spin-transfer
torque, which scales proportional to the current density, dominates
over the torque exerted by the magnetic field caused by the current
flow,which is proportional to the total current.) Dynamical aspects
of the magnetic switching process were addressed in recent
experiments.\cite{kn:kiselev2003,kn:kiselev2004,kn:rippard2004,kn:krivorotov2004}

Over the past few years there has been much theoretical interest in
understanding the spin-transfer torque and its consequences for
hybrid ferromagnet--normal-metal devices. The connection
between spin currents or spin accumulation in the normal metal
spacer layer and the spin torque can be considered 
understood\cite{kn:waintal2000,kn:stiles2002,kn:xia2002}
(see Ref.\ \onlinecite{kn:tserkovnyak2005} for a recent review).
Most calculations
of the response of the magnetization to the spin-transfer
torque have been done in the so-called `macrospin approximation',
assuming that the ferromagnets remain single domains during 
spin-transfer induced switching
events.\cite{kn:sun2000,kn:bazaliy2004,kn:li2004,kn:apalkov2004,kn:kovalev2005,kn:xiao2005} They have 
addressed the precise nature of the magnetic
switching process, the possibility of limit cycles, 
and the temperature
dependence of the spin-transfer torque.
In addition, full micromagnetic simulations have been done
by several groups,\cite{kn:li2002,kn:berkov2005,kn:berkov2005b,kn:torres2005}
{\em e.g.}, to examine
the effect of the Ampere field on the hysteretic switching or
the breakdown of the macrospin model into quasi-chaotic
dynamics at very high current densities. 
While the micromagnetic simulations are
a significant improvement on the macrospin approximation when
it comes accounting for spatial non-uniformities in the switching
process, the existing simulations derive the
spin-transfer torque from an externally fixed spin current, which is
a poor description of the experimental geometries in which the
spin currents are determined as an
intricate combination of spin polarizations caused by all 
ferromagnetic elements in the 
device.\cite{kn:brataas2000b,kn:waintal2000,kn:tserkovnyak2002,kn:tserkovnyak2005}

\begin{figure}[t]
\epsfxsize=0.7\hsize
\epsffile{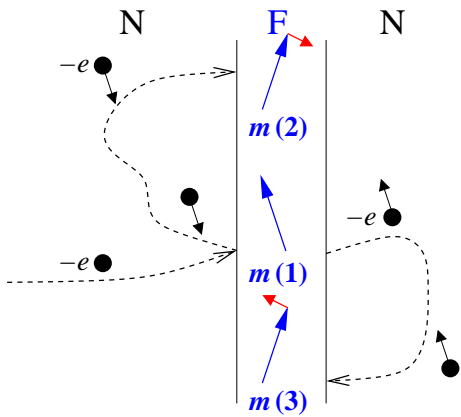}
\caption{\label{fig:cartoon}
Through the spin-transfer torque, an unpolarized electrical
current flowing perpendicular to a thin ferromagnetic layer
can enhance or suppress spin waves. Electrons {\em backscattered}
from the ferromagnet at point $1$ have their spin predominantly 
polarized antiparallel to magnetization direction $\vm(1)$. 
These
electrons exert a torque
on the ferromagnet's magnetization
$\vm(2)$
if they reach the ferromagnet a second time at point $2$, the 
direction of the torque being to enhance an existing spinwave
[{\em i.e.}, to increase any pre-existing
difference between $\vm(1)$ and
$\vm(2)$].
When electrons {\em transmitted} through the ferromagnet reach the
ferromagnet a second time at point $3$, they exert a torque 
that 
suppresses an existing spinwave. If source and drain contacts are not
symmetric, there is a net torque on the ferromagnet, which
enhances or suppresses the spin wave, depending on current
direction.}
\end{figure}

In a recent work, two of the authors showed that a sufficiently
large but unpolarized electrical current flowing perpendicular to
a {\em single} thin ferromagnetic layer can excite spin waves in the
ferromagnet.\cite{kn:polianski2004} These spin waves have 
wavevector perpendicular to
the direction of current flow. 
The key mechanism behind the transverse spin wave instability is 
electron diffusion in the normal-metal contacts perpendicular to the
direction of current flow, see Fig.\ \ref{fig:cartoon}. 
Electrons backscattered from the
ferromagnet are spin polarized, the polarization direction being
antiparallel to the direction of the magnetization at the location
where they were reflected from the ferromagnet. When these electrons
reach the ferromagnetic layer a second time, they typically do 
so at a
different point at the normal-metal--ferromagnet interface. In the
presence of a spin wave, the magnetization direction of the 
ferromagnet will be different at that point, and these electrons will 
transfer the perpendicular component of their
spin to the ferromagnet, thus exerting a spin-transfer torque. The
sign of this torque is to enhance the spin-wave amplitude. A
similar argument can be made for electrons transmitted through the 
ferromagnet, but their torques tend to suppress the spin-wave
amplitude. Typically, source and drain
contacts are asymmetric, and a net spin-transfer torque is
exerted on the ferromagnet. This torque leads to a spin wave
instability for the current direction in which the effect of
backscattered electrons dominates, and not for the other current
direction. Experiments on nanopillars a with single ferromagnetic
layer have verified the
theoretical predictions finding spin-wave instabilities for one 
direction of the current and for asymmetric junctions only.\cite{kn:oezyilmaz2004} 

For a quantitative theory of this transverse spin-wave instability, 
an approach that 
combines a full self-consistent determination of the spin-transfer 
torque and, at the same time, goes beyond the macrospin
approximation is essential.\cite{kn:polianski2004}
Indeed, the macrospin approximation does not allow for non-uniform
spin waves in the ferromagnet, and, whereas an externally imposed
spin transfer torque would predict a similar instability, a
non-self-consistent theory would be quantitatively incorrect
({\em e.g.} predict the wrong wavelength for the spin wave)
because it neglects the coupling between the spin current and the 
spin waves in the ferromagnet.

The possibility of current-induced
non-uniform modes in heterostructures has become of recent interest
in the field, both for single-layer and multilayer 
structures.\cite{kn:ji2003,kn:stiles2004,kn:urazhdin2004,kn:brataas2005,kn:slavin2005,kn:oezyilmaz2005}
In particular, Ji, Chien, and Stiles\cite{kn:ji2003} 
reported experimental and theoretical evidence
that for large ferromagnet thickness, ferromagnet--normal-metal
junctions are unstable to the generation of non-uniform
magnetization modes, but in this case, 
these are longitudinal modes (see also Refs.\
\onlinecite{kn:myers1999} and \onlinecite{kn:chen2004}). 
Further, Stiles, Xiao, and Zangwill
pointed out that transverse spinwaves can be excited even in 
symmetric junctions if the spinwave mode is at not uniform in the
direction of current flow. However, excitation
of these modes requires a higher currents than the transverse spin-waves 
considered here.\cite{kn:stiles2004}

Our previous work,\cite{kn:polianski2004} as well as the other
theoretical works on this and related spin-wave
instabilities,\cite{kn:stiles2004,kn:brataas2005} was a linear stability
analysis, sufficient to predict the onset of the
instability, but not to describe the spin wave amplitude for current
densities larger than the critical current density. Knowledge
of the spin wave amplitude is necessary if one wants to study, {\em
  e.g.,} how the
spin wave instability affects the resistance of the
normal-metal--ferromagnet junction. It is the goal
of this present work to examine in detail the dynamics of the
spin-wave beyond the instability. While we focus on the case of
single-layers, we expect that, in light of the work of
Refs.~\onlinecite{kn:brataas2005,kn:oezyilmaz2005}, 
our qualitative findings will carry over
to the case of tri-layers and heterojunctions.

\begin{figure}[t]
\epsfxsize=0.95\hsize
\epsffile{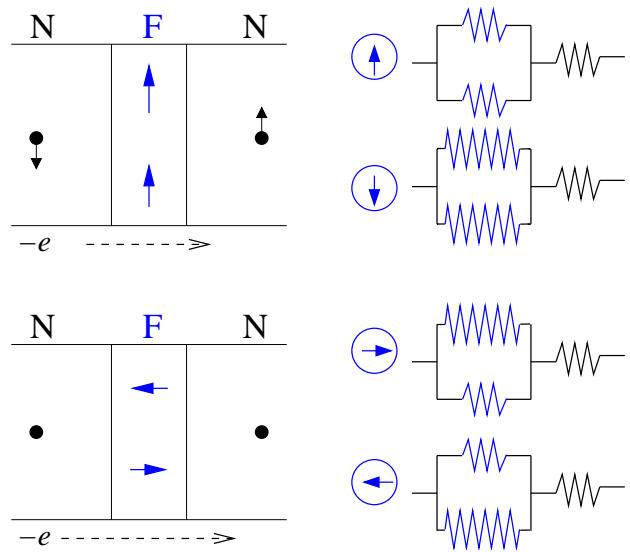}
\caption{\label{fig:cartoon2}
Spin will accumulate in normal metals on both sides of a
ferromagnetic layer with uniform magnetization 
if an unpolarized current is passed through the ferromagnet (top left). A 
large-amplitude spinwave in the ferromagnet reduces the amount of spin 
polarization in the normal-metal regions adjacent to the
ferromagnet and lowers the total resistance of the device
(bottom left). This is shown schematically in the circuit
diagrams (right). The top two circuit diagrams
show the resistances seen by majority and minority
electrons when the magnetization is spatially uniform,
the short and long resistor symbols referring to
minority and majority resistances, respectively.
The ferromagnet with a
large-amplitude spin wave can be seen as a parallel
configuration of ferromagnets with opposite magnetization
directions. The bottom two circuit diagrams show the
resistances seen by two spin directions in this case.
The net resistance is lower in the presence of a
large-amplitude spin wave.}
\end{figure}
Although a quantitative description of how the spinwave
instability affects the resistance of the normal-metal--ferromagnet
junction will be postponed to the next two two sections, the sign of
the effect can be determined using simple considerations.
Once the current density has exceeded the 
critical current density for the spin wave excitation and a
spin wave has been established, the fact that the magnetization
is no longer uniform reduces the amount of
spin accumulation in the normal metal contacts adjacent to the
ferromagnet. A reduction of the spin accumulation in the normal metal 
contacts
causes a reduction of the sample's resistance, see Fig.\ 
\ref{fig:cartoon2} for a schematic
drawing. Indeed, the
experiments of Ref.\ \onlinecite{kn:oezyilmaz2004} observed a small
decrease of the resistance of the nanopillar upon onset of the 
spin-wave instability. The effect of a purely transverse spinwave
instability is opposite to that of a longitudinal
spinwave, which increases the resistance of the
device.\cite{kn:chen2004} The reduction of the spin accumulation
in the normal-metal spacer also lowers the spin-transfer torque, 
thus providing a mechanism to saturate the growth of the spin wave 
amplitude for current densities larger than the critical current
density. Moreover, note that a theory of this effect needs
to combine features of both the micromagnetic approach and
the self-consistent treatment of the spin-transfer torque.

In Sec.\ \ref{sec:2} we consider current densities
slightly above the critical current density. In this regime,
a perturbative treatment in the spin wave amplitude is possible.
In Sec.\ \ref{sec:3} we then perform a detailed numerical simulation 
of a simplified system that allows us to
probe current densities much larger than the critical current
density. Whereas the observed magnetization dynamics in the
presence of a large magnetic field is rather unsurprising ---
there is one stable energy minimum, and the magnetization precesses
around the direction for which energy is minimal ---, in the absence
of an external magnetic field we find a hierarchy of instabilities. 
For very high
currents the system shows chaotic behavior with measurable Lyapunov
exponents.

\section{Perturbative calculation} 
\label{sec:2}

\begin{figure}[t]
\epsfxsize=0.9\hsize
\epsffile{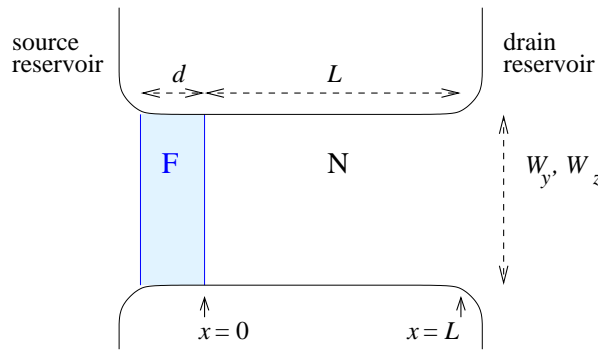}
\caption{\label{Fig:Geometry}
Schematic picture of the normal-metal--ferromagnet--normal-metal
junction considered in our calculations. The ferromagnetic layer 
(F) is 
connected to source and drain reservoirs though normal metal spacers 
(N).
We consider the maximally asymmetric case with only one spacer of
length $L \gg \lsf$.}
\end{figure}

We consider a single ferromagnetic layer, connected to source and
drain reservoirs, see Fig.\ \ref{Fig:Geometry}. 
Between the ferromagnet and the drain reservoir
is a normal-metal spacer, as is common in nanopillar geometries.
There is, however,  no normal-metal spacer between the ferromagnet and the
source reservoir. We use
coordinates $x$, $y$, $z$, where $x$ is the coordinate perpendicular
to the layer structure and $y$ and $z$ are coordinates in the 
plane of the layers.

Both the ferromagnet and the spacer layer have a
rectangular cross section of dimensions $W_y \times W_z$.
The ferromagnet has thickness $d$, which is taken small enough that 
the chemical potential for the conduction electrons and the
the direction $\vm$ of the magnetization of the ferromagnet do not 
depend on the longitudinal coordinate $x$. The normal metal spacer 
has thickness $L$. Transport through the
normal metal spacer is diffusive, with conductivity $\sigma$.

In the normal metal spacer, the charge and spin degrees of the
conduction electrons are described by the equations
\begin{eqnarray}
\nabla^2 \mu_{\rm c} &=& 0, \ \ \ j_x = (\sigma/e)\partial_x \mu_{\rm c}, \nonumber \\
l_{\rm sf}^2 \nabla^2 {\vecmu}_{\rm s} &=& {\vecmu}_{\rm s}, 
\ \ \mathbf{j}_{\rm s} = -(\hbar \sigma/2 e^2) \partial_x {\vecmu}_{\rm s},
\label{Eq:Diff}
\end{eqnarray}
where $\mu_{\rm c}$ and ${\vecmu}_{\rm s}$ are chemical potentials for the
electron density and electron spin respectively, $-e$ is the
electron charge, and $\lsf$ is the spin diffusion length in the
normal metal spacer. Further, $j_x$ is the charge current density and
$\sigma$ is the conductivity of the normal metal leads. The boundary 
conditions for $x=L$ at the drain reservoir is 
\begin{equation}
  \mu_{\rm c} (L) = -eV, \ \ \vecmu_{\rm s}(L) = \mbox{{\boldmath $0$}}.
\end{equation}
Here the argument $L$ refers to the $x$ coordinate. The $y$ and
$z$ coordinates are not written explicitly. 
The second boundary is the interface
between the normal-metal and ferromagnet at $x=0$. 
Since the electron dynamics happens on a time scale that is much
faster than the rate of change of the magnetization direction $\vm$,
this boundary condition can be taken treating $\vm$
in the adiabatic 
approximation,\cite{kn:brataas2000b,kn:tserkovnyak2002}
\begin{eqnarray}
\label{Eq:Boundary}
  j_x (0) &=& \frac{1}{e}
  \left[ g_+ \mu_{\rm c} (0) 
+ g_- {\mathbf m} \cdot \vecmu_{\rm s} (0) 
  \right], \nonumber \\
\mathbf{j}_{\rm s} (0) &=& - \frac{\hbar}{ 2 e^2} 
\left(  g_- \mu_{\rm c} (0) + g_+ \vm \cdot \vecmu_{\rm s} (0) \right) 
\mathbf{m}  \nonumber \\ && \mbox{}
  + \frac{\hbar}{ 2 e^2} g_1 \left( 2 \vecmu_{\rm s} (0) \times \mathbf{m} 
+ \hbar \dot{\vm} \right) \times \mathbf{m} 
 \nonumber \\ && \mbox{}
 +  \frac{\hbar}{ 2 e^2} g_2 \left( 2 \vecmu_{\rm s} (0)  \times \mathbf{m} + 
\hbar \dot{\vm} \right).
\end{eqnarray}
Here $g_{\pm} = (g_{\up} \pm g_{\down})/2$, where $g_{\up}$ and
$g_{\down}$ are interface conductivities for spins aligned parallel 
and anti-parallel to ${\mathbf m}$, and $g_1 + i g_2$ is the 
complex valued `mixing interface conductivity'. The argument ``0''
refers to a coordinate in the normal metal spacer, just outside the
ferromagnetic layer. The charge current and the spin current parallel
to $\vm$ are continuous at the interface. In writing down
Eq.\ (\ref{Eq:Boundary}) we assumed that the two
ferromagnet--normal-metal interfaces are identical, so that
the potentials
$\mu_{\rm c}$ and $\vm \cdot \vecmu_{\rm s}$ drop equally over both interfaces
of the ferromagnet; the component of $\vecmu_{\rm s}$ perpendicular to $\vm$
is zero in the ferromagnet. (It is the non-conservation of spin 
current perpendicular to $\vm$ that gives rise to the spin transfer
torque.) For Co/Cu and Fe/Cr interfaces, these conductivities are
tabulated, see Refs.\ \onlinecite{kn:stiles1996,kn:xia2002}. Typical values are 
in the range $g_2 \ll g_1 \sim g_{\pm} \sim
10^{14} ~ \Omega^{-1} m^{-2}$. For any interface, one has the
constraint $g_1 > g_+ > g_-$.\cite{kn:brataas2000b}

We are interested in the situation in which the magnetization is allowed
to vary in the direction perpendicular to the current flow. 
In this case a large enough
current may cause spin-wave excitations perpendicular to the direction
of current flow.\cite{kn:polianski2004}  
To simplify the notation, we take the 
limit $L \gg l_{\rm sf}$. The spin and charge chemical potentials in
the normal-metal spacer then have the general solution
\begin{eqnarray}
\label{Eq:Ansatz}
  \mu_{\rm c}(\vr) &=& \sum_{\vq} e^{i q_y y + i q_z z}
  a_{\rm c}(\vq) e^{- (q_y^2 + q_z^2)^{1/2} x}
  + \frac{e I x}{W_y W_z \sigma}, \nonumber \\
  \vmu_{\rm s}(\vr) &=& \sum_{\vq}
  e^{i q_y y + i q_z z} e^{- (q_y^2 + q_z^2 + l_{\rm sf}^{-2})^{1/2}
  x} \va_{\rm s}(\vq),
\end{eqnarray}
where $\vq = (0,q_y,q_z)^{\rm T}$ is a wavevector in the $y$-$z$ plane.
The components $q_y$ and $q_z$ take values $q_y = \pi n_y/W_y$,
$q_z = \pi n_z/W_z$ with integers $n_y$ and $n_z$. The Fourier 
expansion coefficients $a_{\rm c}(\vq)$ and $\va_{\rm s}(\vq)$
are real and satisfy
\begin{equation}
  a_{\rm c}(\vq) = a_{\rm c}(-\vq),\ \
  \va_{\rm s}(\vq) = \va_{\rm s}(- \vq).
\end{equation}
We further define the quantities
\begin{eqnarray}
  G_{\rm c}(\vq) &=& (\sigma/2)(q_y^2 + q_z^2)^{1/2}, \\
  G_{\rm s}(\vq) &=& (\sigma/2)(q_y^2 + q_z^2 + l_{\rm sf}^{-2})^{1/2},
\end{eqnarray}
which have the same dimension as the interface conductivities
$g_{\pm}$, $g_1$, and $g_2$. With these definitions, the boundary
condition (\ref{Eq:Boundary}) at the normal-metal--ferromagnet
interface becomes
\begin{widetext}
\begin{eqnarray}
\label{Eq:ToSolve}
  0 &=& 
  - \frac{e I}{W_y W_z} 
  + 2 \sum_{\vq} e^{i q_y y + i q_z z} 
  \left[ G_{\rm c}(\vq) a_{\rm c}(\vq) 
  + g_{\rm +} 
  a_{\rm c}(\vq)
   + g_{\rm -} 
  \va_{\rm s}(\vq) \cdot \vm \right], \nonumber \\
  0 &=&
  \sum_{\vq} e^{i q_y y + i q_z z} \left[ 2 G_{\rm s}(\vq)
  \va_{\rm s}(\vq)
  + (g_{\rm -} a_{\rm c}(\vq) \vm  \right.  
  \left.
 + g_{\rm +} \va_{\rm s}(\vq) \cdot
  \vm) \vm - 2 g_1 (\va_{\rm s}(\vq) \times \vm) \times \vm
   - 2 g_2 \va_{\rm s}(\vq) \times \vm) \right]
  \nonumber \\ && \mbox{}
  - \hbar g_1 \vmd \times \vm - \hbar g_2 \vmd.
\end{eqnarray}
\end{widetext}

Although Eq.\ (\ref{Eq:ToSolve}) gives a set of linear equations
for the expansion coefficients $a_{\rm c}(\vq)$ and $\va_{\rm
  s}(\vq)$, a solution in closed form is not possible for arbitrary
magnetization $\vm(y,z)$. Instead, we expand around the uniform 
equilibrium direction. Hereto we introduce a second coordinate
system with axes labeled $1$, $2$, and $3$, such that $\vm$ 
points along the unit vector $\ve_3$ in the absence of an applied
current, and write 
\begin{equation}
  \vm = m_1 \ve_1 + m_2 \ve_2 + 
(1 - m_1^2 - m_2^2)^{1/2} \ve_3.
\end{equation}
We then perform a Fourier transform, similar to Eq.\
(\ref{Eq:Ansatz})
\begin{equation}
  m_j(y,z) = \sum_{\vq} m_j(\vq) e^{i q_y y + i q_z z},\ \ 
  j=1,2,
\end{equation}
where $m_j(\vq) = m_j(-\vq)$. Finally, expanding in 
powers of $m_1$ and $m_2$, we have solved the
spin and charge chemical potentials to third order in $m_1$ and
$m_2$, which parameterize the deviations from
equilibrium.

In order to complete the calculation, we need to calculate the
rate of change of the magnetization direction $\vm$ in the presence
of the current $I$. Hereto we use the Landau-Lifshitz-Gilbert 
equation,\cite{kn:lifshitz1980,kn:gilbert2004}
\begin{equation}
\vmd = \alpha {\mathbf m} \times \vmd + \vectau_{\rm ex}
  + \vectau_{\rm an}
  + \vectau_{\rm ne},
\label{Eq:LLG1}
\end{equation}
where $\alpha$ is the Gilbert damping coefficient, $\vectau_{\rm ex}$
is the torque arising from exchange, 
$\vectau_{\rm an}$ is the torque from the combined effect of
magnetic anisotropy and an applied magnetic field, and
$\vectau_{\rm ne}$ represents the current-induced spin-transfer
torque. The latter reads\cite{kn:bazaliy1998}
\begin{eqnarray}
\vectau_{\rm ne} &=& \frac{\gamma}{M d}
  (\vj_{\rm s}(0) - \vj_{\rm s}(-d) ) \times \vm) \times \vm
  \nonumber \\ &=& - \frac{\hbar \gamma}{M d e^2}  
\left[ g_1 ( {\vecmu}_{\rm s} \times {\bf m}  
+ \hbar \vmd ) \times {\bf m} \right. \nonumber \\ && \left.
 \mbox{} + g_2 ({\vecmu}_{\rm s} \times {\bf m} + 
\hbar \vmd)\right].
  \label{eq:taune}
\end{eqnarray}    
Here the spin current $\vj_{\rm s}(-d)$ is taken in the
source reservoir, $M$ is the magnetization per unit volume 
and $\gamma$ is the gyromagnetic ratio. Note that the terms 
proportional to the time derivative $\vmd$ 
have contributions from two interfaces while the contribution to 
the torque from the spin chemical potential has a contribution from
the $x=0$ interface only. (All potentials are zero in the source
reservoir.) The exchange torque $\vectau_{\rm ex}$ is
\begin{equation}
  \vectau_{\rm ex} = J \gamma M \nabla^2 \vm \times \vm, 
\end{equation}
where $J$ is the exchange constant.
To linear order in $m_1$ and $m_2$, the
anisotropy torque $\vectau_{\rm an}$ can be written
\begin{eqnarray}
  \vectau_{\rm an} &=&
  - \frac{\gamma}{M} (k_1 m_1 \ve_1 + k_2 m_2 \ve_2) \times \vm,
  \label{eq:tauint}
\end{eqnarray}
where $k_1$ and $k_2$
describe the combined effect of magnetic anisotropy and
an applied magnetic field. 
If anisotropy dominates over
the effect of a magnetic field, higher-order terms in an
expansion in powers of $m_1$ and $m_2$ will be highly
sample specific. Although this case can be
dealt with using the methods presented below, the
result of the calculation has little predictive value if
those coefficients are not known independently. Therefore,
we focus on the opposite limit that the anisotropy term in
Eq.\ (\ref{eq:tauint}) is dominated by magnetic field. Then 
higher-order terms in an expansion in powers of $m_1$ and 
$m_2$ are related to the first-order terms, and one has
\begin{eqnarray}
  \vectau_{\rm an} =
  (k \gamma/M) \ve_3 \times \vm.
  \label{eq:taumagneticfield}
\end{eqnarray}
where we wrote $k_1 = k_2 = k$.
For future reference, we combine the material constants
$J$ and $2k = k_1 + k_2$ into the combinations
\begin{equation}
q_{\rm f}^2 = \frac{k}{J M^2}, 
\mbox{\ \ } j_{\rm f}^2 = 
  \left( \frac{2e}{\hbar}\right)^2 J M^2 k ,
\end{equation}
which have the dimension of inverse length and current density,
respectively. 

We now proceed to report the result of our calculation. The
lowest order result, indicated by a superscript ``$(0)$'', is
\begin{eqnarray} 
  a_{\rm c}^{(0)}(\vq) &=& \frac{ej (g_+ + 2\Go)}{g_{\rm m}(0) g_- }
  \delta_{\vq,0}, \nonumber \\
\va_{\rm s}^{(0)}(\vq) &=& \frac{-e j}{g_{\rm m}(0)} \ve_3 \delta_{\vq,0}.
\end{eqnarray}
Here $j = I/W_y W_z$ is the current density and\cite{kn:polianski2004}
\begin{equation}
  g_{\rm m}(\vq) = \frac{(g_+ + 2\Gq)(g_+ + 2 G_{\rm c}(\vq))}{g_-} - g_-.
\end{equation} 
Writing
$\mu_{\rm c}(L) = -e V = -e (L/\sigma W_y W_z + R) I$, we conclude that the 
resistance $R$ of the ferromagnetic layer is
\begin{equation}
   R = \frac{2}{W_y W_z}
  \frac{\sigma/l_{\rm sf} + g_+}{2 g_+ \sigma/l_{\rm sf} +
  g_+^2 - g_-^2}.
\end{equation}

For the zeroth-order solution, the spin potential $\mu_{\rm s}$ is 
collinear with $\vm$ throughout the sample. Hence, to that order
there is no current-induced torque. This is different when small
deviations from the situation $\vm = \ve_3$ are taken into account
to first order. One finds that the first-order corrections
$a_{\rm c}^{(1)}(\vq)$ and $a_{\rm s3}^{(1)}(\vq)$ are zero. In
order to represent the first-order contributions to the transverse
spin potentials $a_{\rm s1}$ and $a_{\rm s2}$, we use spinor
notation, $a_{\rm s} = (a_{\rm s1},a_{\rm s2})^{\rm T}$ and
$m = (m_1,m_2)^{\rm T}$. Then, defining
\begin{equation}
  D(\vq) = (g_1 + \Gq)^2 + g_2^2,
\end{equation}
we find 
\begin{widetext}
\begin{eqnarray}
  a_{\rm s}^{(1)}(\vq) &=&
  - \frac{e j}{g_{\rm m}(0)} m(\vq)
  + \frac{e j (G_{\rm s}(\vq) - G_{\rm s}(0))}{g_{\rm m}(0) D(\vq)} 
  \left[ (g_1 + \Gq)m(\vq) + i g_2 \sigma_2 m(\vq) \right]
  \nonumber \\ && \mbox{}
  + \frac{\hbar}{2 D(\vq)}
  \left[g_2 G_{\rm s}(\vq) \dot{m}(\vq) + (g_1^2 + g_1 \Gq + g_2^2) 
  i \sigma_2 \dot{m}(\vq) \right].
  \nonumber \\
  \label{eq:afirst}
\end{eqnarray}
\end{widetext}
where $\sigma_2$ is the second Pauli matrix. Note that the
first term on the right hand side is the response to a
uniform rotation of the magnetization, while the second and third
terms give the response to a non-uniform and time-dependent
magnetization.  

The potentials are substituted into Eq.\ (\ref{eq:taune}) to find
the current-induced torque, and then into the Landau-Lifshitz-Gilbert 
equation (\ref{Eq:LLG1}) to find the rate of change of the
magnetization. The current-induced torque has contributions
proportional to the time derivative $\vmd$, which lead to a 
renormalization of the Gilbert damping parameter $\alpha$ and
the the gyromagnetic ratio $\gamma$. 
The renormalized Gilbert damping parameter $\tilde
\alpha$ and gyromagnetic ratio $\tilde \gamma = 
\gamma/\tilde \beta$ depend on the
transverse wavevector $\vq$ and read
\begin{eqnarray}
\tilde\alpha &=& \alpha + \frac{\hbar^2 \gamma (g_1 + \Gq)}{2 M d e^2}
\left[ 1 - \frac{\Gq^2}{D(\vq)} \right], \nonumber \\
\tilde\beta &=& 1 + \frac{\hbar^2 \gamma g_2}{2 M d e^2} 
\left[ 1 + \frac{\Gq^2}{D(\vq)} \right].
  \label{eq:aren}
\end{eqnarray}
In the 
macrospin limit $q \rightarrow 0$, these modifications coincide with
the renormalized values originally
reported in Ref.~\onlinecite{kn:tserkovnyak2002}. 

Again using two-component spinor notation, the
complete Landau-Lifshitz-Gilbert equation then becomes
\begin{eqnarray}
\label{Eq:LLGM}
  (\tilde \beta \openone_2 + i \sigma_2 \tilde \alpha)
  \dot{m}(\vq) =
  A(\vq) m(\vq),
\end{eqnarray}
with
\begin{eqnarray}
  A(\vq) &=& \tau_{\parallel}^{(1)} \openone_2 - i \sigma_2
  \left[
  \tau_{\perp}^{(1)} 
  + \frac{\hbar \gamma j_{\rm f}(q^2 + q_{\rm f}^2)}{2 e
  q_{\rm f} M} \right]
  \nonumber \\ && \mbox{}
  + \sigma_3 \frac{\gamma (k_1 - k_2)}{2 M}
\end{eqnarray}
and
\begin{eqnarray}
  \tau_{\parallel}^{(1)}(\vq) &=&
\frac{\hbar \gamma e j}{M d e^2} \frac{g_1^2 + g_2^2 + g_1 \Gq }
{g_{\rm m}(0) D(\vq)}  [\Gq - \Go] , \nonumber \\
  \tau_{\perp}^{(1)}(\vq) &=&
\frac{\hbar \gamma e j}{M d e^2} \frac{g_2 \Gq}{g_{\rm m}(0) D(\vq)} [\Gq - \Go].
\end{eqnarray}

In the absence of a current, any spatial modulation of the 
magnetization is damped. However, a sufficiently large positive
current $I$ can overcome the damping, and cause a spatial modulation
of $\vm$ to grow in time, rather than decay. 
(A positive current $I$ corresponds to electron
flow in the negative $x$ direction.)
The instability 
condition is easily obtained from Eq.\ (\ref{Eq:LLGM})
\begin{equation}
 \tau_{\parallel}^{(1)}(\vq) \frac{ \tilde\beta(\vq) }{\tilde\alpha(\vq)}
  > \frac{\hbar \gamma j_{\rm f}(q^2 + q_{\rm f}^2)}{2 e q_{\rm f} M}
  + \tau_{\perp}^{(1)}(\vq).
  \label{eq:critcond}
\end{equation}
We can analyze this result in different limits.
For a ferromagnetic layer with sufficiently small transverse
dimensions, $W_{y}, W_{z} \lesssim (l_{\rm sf}/q_{\rm f}^2)^{1/3}$
if $l_{\rm sf} q_{\rm f} \gg 1$,
the instability happens at wavevector $\vq =
(\pi/W_y) \hat y$ or $\vq = (\pi/W_z) \hat z$, whichever is 
smallest, and the critical
current follows directly from Eq.\ (\ref{eq:critcond}). For
wider layers, the critical current density $j_{\rm c}$ and critical 
wavevector $\vq_{\rm c}$ are found as the current-density wavevector 
pair for 
which the onset of the instability condition happens at the lowest 
current density. 
 
This condition can be simplified in the limit of a very
thin ferromagnetic layer, $d \to 0$, neglecting terms proportional
to $g_2$ (which is numerically smaller than $g_1$), and
for wavenumbers $q \ll q_{\rm f}$. We then find
that the critical current follows from minimizing the relation
\begin{eqnarray}
  j_{\rm c}(\vq)
   &=& \frac{\hbar^2 \gamma g_{\rm m}(0) j_{\rm f}}{2 M q_{\rm f} e^2} \frac{ q^2 + q_{\rm f}^2}
{1 - (1 + q^2 l_{\rm sf}^2)^{-1/2}}.
   \label{eq:jcsimple}
\end{eqnarray}
In the limit $l_{\rm sf} \gg 1/q_{\rm f}$, this gives\cite{kn:polianski2004}
\begin{equation}
  q_{\rm c} = (q_{\rm f}^2/2 l_{\rm sf})^{1/3},\ \
  j_{\rm c} = \frac{\hbar^2 \gamma g_{\rm m}(0) q_{\rm f} j_{\rm f}}{2 M e^2}.
  \label{eq:jcresult}
\end{equation}
(The result for $j_{\rm c}$ was reported incorrectly in 
Ref.\ \onlinecite{kn:polianski2004}. Note that the condition
$q_{\rm c} \ll q_{\rm f}$, which was used to derive Eq.\ (\ref{eq:jcsimple})
is consistent with Eq.\ (\ref{eq:jcresult}) if $l_{\rm sf}
\gg 1/q_{\rm f}$.)
Note that $q_{\rm f}$ increases with an applied magnetic field, so that
this limit becomes relevant even for the case of a normal metal
with strong spin relaxation if the magnetic field is large
enough.
In the limit $l_{\rm sf} \ll 1/q_{\rm f}$ of strong spin relaxation
and weak anisotropy, one has
\begin{equation}
  q_{\rm c} = (4/3)^{1/4} (q_{\rm f}/l_{\rm sf})^{1/2},\ \
  j_{\rm c} = \frac{\hbar^2 \gamma g_{\rm m}(0) j_{\rm f}}{M q_{\rm f} 
  \lsf^2 e^2}.
\end{equation}

At the
critical current density, the trajectory of the magnetization is
a simple ellipse (circle in the case of large magnetic fields).
The ellipse is described by the coordinate transformation
$ m_1 = r ( \cos \theta \cos \phi + 
  \eta \sin \theta \sin \phi ),$ and $m_2 =  r ( \sin \theta \cos \phi -
  \eta \cos \theta \sin \phi ).$  The solution of the magnetization
dynamics at the critical current then gives $\phi = \omega_0 t$ 
and $r$ constant, 
where $\omega_0^2 = \omega_+^2 - \omega_-^2$, $\eta = 
(\omega_+ - \omega_-)/\omega_0$ and
\begin{eqnarray}
  \label{eq:omega}
\omega_+^{-1} &=& 
  \frac{2 M c e^2 q_{\rm f} \cos (2 \theta)}{\gamma (q_{\rm f}^2 + q_{\rm c}^2) j_{\rm f}}
  \nonumber \\ && \mbox{} -
  \frac{2 M c e^2 q_{\rm f} g_2 G_{\rm s}(\vq) \sin(2 \theta)}
  {\gamma  j_{\rm f} (q_{\rm f}^2 + q_{\rm c}^2) (g_1^2 + g_2^2 + g_1 \Gq) )}, \\
\omega_-^{-1} &=&
  \frac{2 M c \hbar}{\gamma(k_1-k_2)}.
\end{eqnarray}
and $c, \theta$ are obtained from $\tilde\alpha = c \sin 2 \theta,
\tilde\beta = c \cos 2 \theta$.  For the case of a large applied
magnetic field, $k_1 = k_2 = k$, and neglecting
$g_2$, we have $\eta = 1$ and 
\begin{equation}
  \omega_0 = 
  \hbar \gamma j_{\rm f} (q_{\rm c}^2 + q_{\rm f}^2)/(2 e q_{\rm f} M).
\end{equation}
Note that, although the applied
current has a large effect on the stability of the ellipsoidal
motion (precession is damped for $j < j_{\rm c}$ and unstable for
$j > j_{\rm c}$), its effect on the precession frequency is small.
To a good approximation, the precession frequency equals the
ferromagnetic resonance frequency in the absence of a current.

Whereas the first-order calculation allows one to find the current
density at which the spin-wave instability sets in and the angular
form of the low-amplitude excitations, it does not provide information
about the magnitude of the spinwave oscillation for $j > j_{\rm c}$,
or about the effect of the spinwave oscillation on the resistance of
the ferromagnetic layer.
This information can only be obtained from the analysis of the
magnetization dynamics beyond first-order in the amplitude. Such
a program proceeds along the same lines as the first-order calculation
shown above: Calculation of the potentials for charge and spin in the
presence of a non-uniform and time-dependent magnetization, followed
by a calculation of the current-induced torque and the rate of change
of the magnetization. We have carried out this program to third order 
in $m_1$ and $m_2$, and list some of our general results in the 
appendix. However, as this calculation involves higher-order
contributions to the anisotropy torque $\vectau_{\rm an}$, 
for which the expansion constants are unknown, we find that
this calculation has little predictive value. Instead, we focus on 
the limit in which all magnetic anisotropy arises from an applied
magnetic field. In this limit, $\vectau_{\rm an}$ is known, 
cf.\ Eq.\ (\ref{eq:taumagneticfield}), and a theoretical
analysis is useful.      

An important simplification is that the higher-order analysis is 
necessary for the Fourier components $m_1(\vq_{\rm c})$ and 
$m_2(\vq_{\rm c})$ at the critical wavevector only. Hence, we
need to consider only a single Fourier component in our considerations
below. Solving for the leading (second order) correction to the
charge potential, we find an expression that depends on the
magnetization
amplitude, to second order in $m_1$ and $m_2$, and on the
time derivatives. Only first-order time-derivatives appear, which
can be eliminated using the Landau-Lifshitz-Gilbert equation
(\ref{Eq:LLGM}). For the case of a large applied magnetic field,
the magnetization precession is circular, and one has
\begin{equation}
  m_1(\vq_{\rm c}) \dot m_2(\vq_{\rm c}) - m_2(\vq_{\rm c})
\dot m_1(\vq_{\rm c}) = \omega_0 r(\vq_{\rm c})^2,
  \label{eq:momega}
\end{equation}
where we abbreviated
\begin{equation}
  r(\vq_{\rm c})^2 = m_1(\vq_{\rm c})^2 + m_2(\vq_{\rm c})^2.
\end{equation}
The precession frequency $\omega_0$ given by Eq.\ (\ref{eq:omega}) 
above. We then find
\begin{eqnarray}
  \lefteqn{a_{\rm c}^{(2)}(0) =
  \frac{2 (G_{\rm s}(0) - G_{\rm s}(\vq_{\rm c})) r(\vq_{\rm c})^2 }
  {D(\vq_{\rm c}) g_{\rm m}(0)^2}}
  \nonumber \\ && \mbox{} \times
  \left[ \omega_0 g_{\rm m}(0)
  \left(D(\vq_{\rm c}) - G_{\rm s}(\vq_{\rm c}) (g_1 + G_{\rm s}(\vq_{\rm
  c}))\right)
  \right. \nonumber \\ && \left. \mbox{} 
  - 2 e j_{\rm c}
  \left(D(\vq_{\rm c}) +  (G_{\rm s}(0) - G_{\rm s}(\vq_{\rm c})) (g_1 +
  G_{\rm s}(\vq_{\rm c})) \right)\right] \nonumber \\
\end{eqnarray}

Solving for the leading (third) order torque, we note that the
third order torque depends not only on the magnetization
amplitudes $m_1(\vq_{\rm c})$ and $m_2(\vq_{\rm c})$, but also
on their time derivative $\dot{m}_1$ and $\dot{m}_2$. The 
time derivatives appear to first, second, and third order
in the expansion. The dependence on $\dot{m}^{(3)}$ leads to
the same modifications to the Gilbert damping and gyromagnetic 
ratio as for the first-order current-induced torque calculated above. 
The dependence on $\dot{m}^{(2)}$ is through the $3$-component
only, which can be written as
\begin{equation}
  \dot{m}_3^{(2)} = - m_1 \dot{m}_1^{(1)} -
m_2 \dot{m}_2^{(1)}.
\end{equation}
The first-order time derivatives
$\dot{m}^{(1)}$ can be expressed in terms of $m_1$ and $m_2$ 
using Eq.\ (\ref{eq:momega}) [or, in the general case, using
Eq.\ (\ref{Eq:LLGM})]. For the anisotropy torque $\tau_{\rm an}$ we take
the contribution from the magnetic field only. Hence,
\begin{eqnarray}
  \tau_{\rm ex}(\vq_{\rm c}) +
  \tau_{\rm an}(\vq_{\rm c}) &=&
  \frac{\hbar \gamma j_{\rm f}}{2 e q_{\rm f} M}
  \left[ q_{\rm c}^2 + q_{\rm f}^2 + \frac{q_{\rm c}^2}{2} r(\vq_{\rm c})^2
  \right]
  \nonumber \\ && \mbox{} \times
   (-i \sigma_2) m(\vq_{\rm c}).
\end{eqnarray}
Thus proceeding, we find that the third-order equation for
the rate of change of the magnetization direction reads
\begin{eqnarray}
\label{Eq:LLGM3}
  (\tilde \beta \openone_2 + i \sigma_2 \tilde \alpha)
  \dot{m}(\vq)^{(3)} =
  A(\vq)^{(3)} m(\vq)^{(3)},
\end{eqnarray}
with
\begin{widetext}
\begin{eqnarray}
  A(\vq_{\rm c})^{(3)} &=&
  - \frac{1}{2} r(\vq_{\rm c})^2
  \left[2\tau_{\parallel}^{(3)}(0) \openone_2
  - 2 i \sigma_2 \tau_{\perp}^{(3)}(0)
  + \tau_{\parallel}^{(3)}(2 \vq_{\rm c}) \openone_2
  - i \sigma_2 \tau_{\perp}^{(3)}(2 \vq_{\rm c})
  + 3 \tilde{\alpha} \omega_0  \openone_2 +
  \frac{\hbar \gamma j_{\rm f} q_{\rm c}^2}{2 e q_{\rm f} M} i \sigma_2
  \right],
\end{eqnarray}
and
\begin{subequations}
\begin{eqnarray}
  \tau^{(3)}_{\parallel}(\vk) &=& \frac{\hbar \gamma g_1}{Md e^2(g_1 + G_\vq)} \left\{
\frac{[g_+ + 2 G_{\rm c}(\vk)] [ g_{+} + 2 \Gq]  - g_-^2 }
     {[g_+ + 2 G_{\rm c}(\vk)] [ g_{+} + 2 \Gk]  - g_-^2 }
\right\}
  \nonumber \\ && \mbox{} \times
\left\{ \frac{ej}{g_{\rm m}(0)} [\Gk - \Go] 
  +  \left[ g_1 \hbar \omega_0 - \frac{2 ej (g_1 + \Go)}{g_{\rm m}(0)}
\right] \frac{\Gk - \Gq}{g_1 + \Gq} \right\}, \\
  \tau^{(3)}_{\perp}(\vk) &=&  \frac{ \hbar \gamma g_2 \, \Gq [\Gk
  - \Gq] }{Md e^2 [g_1 + \Gq]^3}
 \left\{
\frac{[g_+ + 2 G_{\rm c}(\vk)] [ g_{+} + 2 \Gq] - g_-^2 }
     {[g_+ + 2 G_{\rm c}(\vk)] [ g_{+} + 2 \Gk] - g_-^2 }
\left[ g_1 \hbar \omega_0 -\frac{ej (g_1 + \Go) }{g_{\rm m}(0)} \right] \right. \nonumber \\
&& \left. \mbox{\hspace{1.5in}}
+  \frac{[(g_+ - 2 g_1) (2 G_{\rm c}(\vk) + g_{+}) - g_-^2][\Gq - \Go] ej}
    {g_{\rm m}(0)[(g_+ + 2 G_{\rm c}(\vk)) ( g_{+} + 2 G_{\rm s}(\vk)) 
  - g_-^2]}
  \right\}.
\end{eqnarray}
\end{subequations}

Solving the differential equation for $m$, one finds that the 
precession amplitude for current density $j$ slightly above the
critical current density $j_{\rm c}$ reads
\begin{eqnarray}
  r(\vq_{\rm c})^2 &=&
  \frac{\hbar \gamma j_{\rm f} (j -j_{\rm c})}{ e q_{\rm f} M j_{\rm c}}
  \frac{\tilde \alpha (q_{\rm c}^2 + q_{\rm f}^2)}{ \tilde \beta
  \left[ 2 \tau^{(3)}_{\parallel}(0) +
  \tau^{(3)}_{\parallel}(2 \vq_{\rm c}) -
  3 \tau^{(1)}_{\parallel}(\vq_{\rm c}) \right] - 
 \tilde \alpha \left[
  2 \tau^{(3)}_{\perp}(0) +
  \tau^{(3)}_{\perp}(2 \vq_{\rm c}) -
  \hbar \gamma j_{\rm f} q_{\rm c}^2/(2 e q_{\rm f} M) \right]}.
\label{eq:amplitude}
\end{eqnarray}
\end{widetext}
The result takes a simpler form in the limit $g_2 \to 0$ (since
$g_2$ is numerically smaller than $g_1$),
$d \to 0$, and $1/l_{\rm sf} \ll q_{\rm c} \ll q_{\rm f}$,
\begin{eqnarray}
  r(\vq_{\rm c})^2 &=&
  \frac{(j - j_{\rm c}) ( g_+^2  - g_-^2 +  g_+ \sigma/l_{\rm sf} )}
  {j_{\rm c}(  2 g_1 g_+ +  g_+ \sigma/l_{\rm sf} - g_+^2 + g_-^2)}.
\end{eqnarray}
Since $g_1 > g_+ > g_-$ we conclude that the $r(\vq_{\rm c})^2 > 0$
is positive if $j > j_{\rm c}$, which excludes hysteretic
behavior.

In the same limit we can also calculate the change in
frequency of the spinwave given by  
\begin{eqnarray}
\frac{\omega}{\omega_0} =  1 + \frac{ q_{\rm c}^2 r(\vq_{\rm c})^2}
  {3(q_{\rm c}^2 + q_{\rm f}^2)}.
\end{eqnarray}
Since the prefactor of the second term is much smaller than unity,
$q_{\rm c} \ll q_{\rm f}$ for the parameter regime of interest, we
conclude that in the regime of perturbation theory, there is hardly any
change from the ferromagnetic resonance frequency. 

Finally, at the onset of the 
spin-wave instability, the resistance of the ferromagnetic layer
acquires a small negative correction
\begin{eqnarray}
 \frac{R}{R_0} &=& 1
 + 
  \frac{a_{\rm c}^{(2)}(0)}{a_{\rm c}^{(0)}(0)}, \nonumber \\
&\approx& 1 - 
\frac{2 (\sigma/l_{\rm sf} + 3 g_1)  g_-^2  r(\vq_{\rm c})^2}{
(\sigma/l_{\rm sf} + g_+ )( g_+^2  - g_-^2 + g_+ \sigma/l_{\rm sf} )}.
  \label{eq:Rchange}
\end{eqnarray}
(In the second line we took the limits $g_2 \to 0$, $d \to 0$,
and used $1/\lsf \ll q_{\rm c} \ll q_{\rm f}$.)
This resistance decrease is anticipated on physical grounds
since the non-uniform mode allows for an increased transmission 
of minority elections that diffuse along the transverse direction 
--- see Fig.\ \ref{fig:cartoon2} and the corresponding 
discussion in Sec.\ \ref{sec:1}.

\section{Numerical Calculation}
\label{sec:3}

The calculations in the preceding section are valid for currents
close to the onset of the instability. For
currents much larger than the critical current, we need to go beyond
perturbation theory to obtain the dynamics. Hereto we numerically
solve for the magnetization dynamics and its effect on the resistance
of the ferromagnetic layer.

\begin{figure}[t]
\epsfxsize=0.5\hsize
\hspace{0.05\hsize}
\epsffile{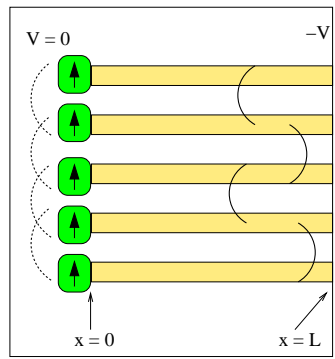}
\caption{\label{Fig:Numerical_Model}
Schematic drawing of the model solved numerically. The continuous
magnet is replaced by $N$ magnets (left), each coupled to a normal-metal
wire (right). The wires are coupled via transverse diffusion 
(shown schematically as solid lines); the magnets are coupled via
the exchange interaction (shown schematically as dashed lines).}
\end{figure}

In our numerical analysis, we assume $W_z \ll W_y$ and impose that
the magnetization direction $\vm(y,z)$ does not depend on $z$.
The remaining two-dimensional problem is replaced by a finite number 
of one-dimensional problems by substituting the normal-metal spacer
and the ferromagnetic layer by $N$ normal metal channels, each 
attached to a magnet with magnetization direction $\vm(n)$, 
$n=1, \ldots, N$. In order to model a higher-dimensional structure,
electrons are allowed to diffuse between the channels, whereas the
$N$ magnets interact via an exchange energy. A schematic drawing 
of this 
model is shown in Fig.\ \ref{Fig:Numerical_Model}. 

In this discretized model, the potentials for charge and spin
obey the equations
\begin{eqnarray}
\partial^2_x \mu_{\rm c}(n,x) &+& \left( \frac{N}{W_y} \right)^2 [ \mu_{\rm c}(n+1, x)
+ \mu_{\rm c}(n-1, x) \nonumber \\ 
&& \mbox{} - 2 \mu_{\rm c}(n, x)] = 0, \nonumber \\ 
\partial^2_x \vecmu_{\rm s}(n,x) &+& \left( \frac{N}{W_y} \right)^2 [ \vecmu_{\rm s}(n+1, x) +
\vecmu_{\rm s}(n-1, x) \nonumber \\
&& \mbox{} - 2 \vecmu_{\rm s}(n, x)] = \frac{\vecmu_{\rm s}(n, x) }{
\lsf^2}. 
  \label{eq:dmu}
\end{eqnarray}
Equations for the boundary channels, $n=1$ and $n=N$, are obtained
by setting $\mu_{\rm c,s}(0,x) = \mu_{\rm c,s}(1,x)$ and $\mu_{\rm c,s}(N+1,x) 
= \mu_{\rm c,x}(N,x)$. 
The general solution of Eq.\ (\ref{eq:dmu}) is of the form
\begin{eqnarray}
  \mu_{\rm c}(n,x) &=& 
  2 \sum_{l=0}^{N-1} a_{\rm c}(l) \cos \left[ l \pi (n + 1/2)/N \right]
  e^{-q_{\rm c}(l) x}
  \nonumber \\ && \mbox{} + \frac{e I x}{\sigma W_y W_z}.
  \nonumber \\
  \vecmu_{\rm s}(n,x) &=&
  2 \sum_{l=0}^{N-1} \va_{\rm s}(l) \cos \left[ l \pi (n + 1/2)/N \right] 
  e^{-q_{\rm s}(l) x}, \nonumber \\
\end{eqnarray}
with
\begin{eqnarray}
  q_{\rm c}(l)^2 &=&
    4 (N/W_y)^2  \sin^2 \left( l \pi/2 N) \right), \nonumber \\
  q_{\rm s}(l)^2 &=& l_{\rm sf}^{-2} +
    4 (N/W_y)^2  \sin^2 \left( l \pi/2 N) \right).
\end{eqnarray}
The boundary conditions at $x = 0$ (normal-metal--ferromagnet
interface) are given by Eq.\ (\ref{Eq:Boundary}).

The magnetization dynamics
is given by the Landau-Lifshitz-Gilbert equation (\ref{Eq:LLG1}),
with a discretized exchange torque $\vectau_{\rm ex}$,
\begin{eqnarray}
  \vectau_{\rm ex}(n) &=&
  \frac{J \gamma M N^2}{W_y^2}
  \left[\vm(n+1) + \vm(n-1) \right] \times \vm(n), 
  \nonumber \\
\end{eqnarray}
For the anisotropy torque we consider two different cases:
The limit of a large applied magnetic field,
\begin{equation}
  \vectau_{\rm an} = \frac{k \gamma}{M} \ve_3 \times \vm(n),
  \label{eq:taueqnum1}
\end{equation}
as well as the case of no applied field, where we take a
simple model for the torque arising from magnetocrystalline 
and shape anisotropy,
\begin{eqnarray}
  \vectau_{\rm an}(n) &=&
  - \frac{\gamma}{M} [k_1 m_1(n) \ve_1 + k_2 m_2(n) \ve_2] 
  \times \vm(n). \nonumber \\
  \label{eq:taueqnum2}
\end{eqnarray}

The Landau-Lifshitz-Gilbert equation, together with the
boundary conditions at $x=0$, are sufficient to determine
the $4N$ expansion coefficients $a_{\rm c}(l)$ and
$\va_{\rm s}(l)$, $l=0,\ldots,N-1$, and the time derivative
of the magnetization directions $\vm(n)$, $n=1,\ldots,N$.
Our numerical procedure consists of first expressing $\dot{\vm}(n)$
in terms of the potential expansion coefficients $a_{\rm c}(l)$ and
$\va_{\rm s}(l)$ using the Landau-Lifshitz-Gilbert equation,
and then solving for the potential expansion coefficients using
the boundary condition at $x=0$. 

For the practical implementation of this scheme, it is useful
to define $3 \times 3$ matrices ${\cal M}$ and 
${\cal R} = \vm \vm^{T}$ such that for any vector ${\bf v}$, 
${\bf v} \times \vm = {\cal M}
 {\bf v}$ and ${\cal R} - \openone_3 = {\cal M}^2$. In $3 \times 3$
matrix notation, the time derivative of the magnetization vector
can be expressed in terms of the potential coefficients as 
\begin{widetext}
\begin{eqnarray}
\dot{\vm}(n) &=& 
  \frac{\beta' \openone_3 + \alpha'^2 {\cal R}/\beta' - 
  \alpha' {\cal M}}{\alpha'^2 + \beta'^2}
  \nonumber \\ && \mbox{} \times
  \left\{ {\cal M} [\vectau_{\rm ex} + \vectau_{\rm an}]
  + \frac{2 \hbar \gamma}{Mde^2} \sum_{l=0}^{N-1} 
  [{\cal M} g_1 + g_2 \openone_3] {\cal M}
  \va_{\rm s}(l) \cos \left[ l \pi (n + 1/2)/N \right]
  \right\},
\end{eqnarray}
where $\alpha' = \alpha + \hbar^2 \gamma g_1 /(Mde^2)$ 
and $\beta' = 1 + \hbar^2 \gamma g_2/(Mde^2)$. In turn, the potential
coefficients $\va_{\rm s}(l)$ are obtained from inverting a
$4N$ dimensional matrix equation,
\begin{eqnarray}
\sum_{l = 0}^{N-1} 
2 \cos \left( l \pi (n + 1/2)/N \right)
\left[ \begin{array}{cc}
      2 \sigma q_{\rm c}(l) + 2 g_+  & 
	2 \vm^{\rm T} g_-   
      \\       2 \vm g_-  & 
      \sigma q_{\rm s}(l)  \openone_3 + 4  \chi_1 \end{array}
      \right]  \left( \begin{array}{c} a_{\rm c}(l) \\ {\mathbf a}_{\rm s}(l)
\end{array} \right) 
  = \frac{2 e I}{W_y W_z}
  \left( \begin{array}{c} g_+ \\ g_- \vm \end{array} \right)
  + \left( \begin{array}{c} 0 \\ \vecchi_2 \end{array} \right),
\end{eqnarray}
where we abbreviated
\begin{eqnarray}
\chi_1(n) &=& g_- {\cal R} 
          - {\cal M} (g_1 {\cal M} + g_2 \openone_3)
  \nonumber \\ && \mbox{} 
           + \frac{\hbar^2 \gamma}{2 M d e^2(\alpha'^2+\beta'^2)}
   (g_1 {\cal M} + g_2 \openone_3)
	    [\beta' \openone_3 + \alpha'^2 {\cal R}/\beta' - 
  \alpha' {\cal M}] {\cal M}
	   (g_1 {\cal M} + g_2 \openone_3), \\
\vecchi_2(n) &=& \frac{2 (g_1 {\cal M} + g_2 \openone_3) \hbar^2 \gamma }
 {M d e^2 (\alpha'^2+\beta'^2)}
  [\beta' \openone_3 + \alpha'^2 {\cal R}/\beta' - 
  \alpha' {\cal M}] {\cal M} [\vtau_{\rm ex}(n)+ \vtau_{\rm an}(n)].
\end{eqnarray}
\end{widetext}

We have performed numerical simulations for $N$ ranging between
$10$ and $20$, although all data shown are for $N=10$ and $N=11$.
We verified that there is
no qualitative dependence on the parity of $N$ in our simulations.
A small random
torque was added at each time step to mimic the effect of a 
small but finite temperature. (The corresponding 
temperature obtained from the fluctuation-dissipation theorem was
less then a mK.\cite{kn:apalkov2004})

Below we present our results. We first consider the case in
which the anisotropy torque is dominated by an applied magnetic
field, taking Eq.\ (\ref{eq:taueqnum1}) for the anisotropy torque
$\vectau_{\rm an}$. We
then consider the case in which there is no applied magnetic
field, taking Eq.\ (\ref{eq:taueqnum2}) for $\vectau_{\rm an}$.
The latter case is qualitatively different from the former,
as it has two stable equilibria for $\vm$ ($\vm = \ve_3$ and
$\vm = -\ve_3$), whereas in the presence of a large applied field
the equilibrium position is at $\vm = \ve_3$.

\subsection{Large applied magnetic field}

\begin{figure}[t]
\epsfxsize=1\hsize
\hspace{0.05\hsize}
\epsffile{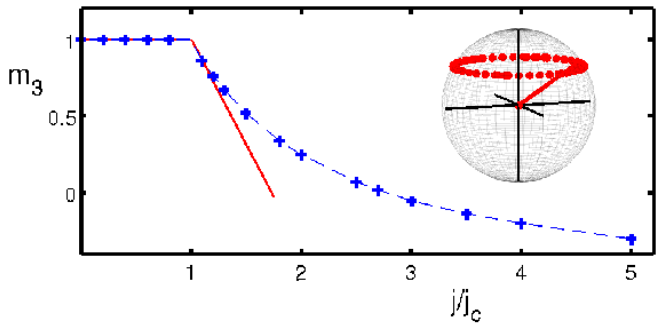}
\caption{\label{Fig:MagField}
Main panel shows the magnetization component $m_3(1)$ of the first 
magnet, as a function of applied current.  The solid line is obtained from
the perturbation
theory result (\protect\ref{eq:amplitude}), while the dashed line is a
guide to the eye.  In a large magnetic field, the motion is circular.
an example is shown in the inset where $j=1.5 j_{\rm c}$.}
\end{figure}

For the numerical simulations with a magnetic field, we
took values for the various parameters as follows:
thickness $d=0.2~\mathrm{nm}$, Width $W_y = 55~\mathrm{nm}$, as
is appropriate for typical nanopillar experiments,\cite{kn:katine2000}
spin-diffusion length
$\lsf = 100~\mathrm{nm}$, $\sigma/\lsf = 10^{15} \Omega^{-1} 
\mathrm{m}^{-2}$, $g_1 = 5.5 \times
10^{14}~\Omega^{-1} \mathrm{m}^{-2}$, $g_2 = 0.3 \times
10^{14}~\Omega^{-1} \mathrm{m}^{-2}$, $g_{\up} = g_+ + g_- = 4.2 \times
10^{14}~\Omega^{-1} \mathrm{m}^{-2}$, $g_{\down} = g_+ - g_- = 3.3 \times
10^{14}~\Omega^{-1} \mathrm{m}^{-2}$. The interface conductivities
are taken from numerical calculations for a disordered Cu/Co
interface;\cite{kn:xia2002} the conductivity $\sigma$ and the spin
relaxation
length $\lsf$ are consistent with those in Cu. We further took
$\alpha = 0.01$, $\hbar \gamma g_1/M d e^2 = 0.0138$, $j_{\rm f} = 10^{12} \mathrm{A}/\mathrm{m}^2$,
$q_{\rm f} = 10^{-1} \mathrm{nm}^{-1}$ (as is 
appropriate for Co, see Ref.\ \onlinecite{kn:wohlfahrt1980}; the 
magnetic field corresponding to the values of $j_{\rm f}$ and $q_{\rm f}$ listed
above is of a strength comparable to the intrinsic anisotropy
energy). For these parameters, the width of the sample
is so small that the spinwave wavenumber $q$ is set by the finite
sample width, $q = \pi/W_y$.

\begin{figure}[t]
\epsfxsize=1\hsize
\hspace{0.05\hsize}
\epsffile{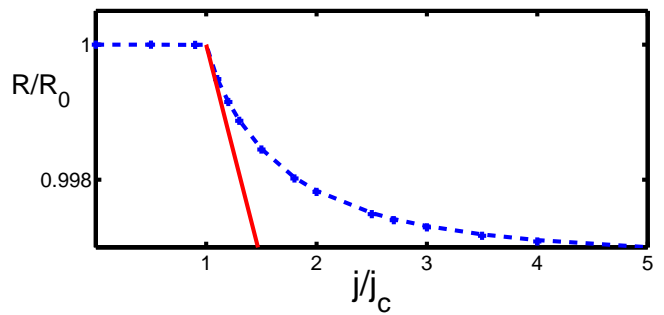}
\caption{\label{Fig:RMagField}
Resistance of the ferromagnetic layer, as a function of applied
current (crosses). The solid line is obtained from
the perturbation
theory result (\protect\ref{eq:Rchange}), while the dashed line is a
guide to the eye.}
\end{figure}

For current densities below $j_{\rm c}$, no spinwaves are excited. Simulation runs
in which the magnetization is tilted away from the easy axis
$\ve_3$ show damped precession towards the equilibrium magnetization
direction $\vm = \ve_3$. For current densities above $j_{\rm c}$,
a spin-wave with wavenumber $q=\pi/W_y$ is excited. Each magnet
$n$ in our simulation $n=1,\ldots,N$ shows circular precession 
around the direction of the applied magnetic field, see Fig.\
\ref{Fig:MagField}, inset. The amplitude of the oscillation increases 
with current as
predicted by the perturbation theory of the preceding section. The
$3$-component of the magnetization is a constant of the motion 
and can be monitored to measure the amplitude.
Numerical results for $m_3$ for the magnet $n=1$ are shown in 
Fig.~\ref{Fig:MagField} as a function of current density,
together with a comparison of our 
numerical results with the perturbative result (\ref{eq:amplitude}).
With a large applied field, the magnetization dynamics
remains regular even for current densities much larger than $j_{\rm
  c}$. The effect of the spin-wave instability on the resistance
of the ferromagnetic layer is shown in Fig.\ \ref{Fig:RMagField}.

\subsection{No applied magnetic field}

We have also performed numerical simulations in the absence
of an applied magnetic field. Hereto, we choose Eq.\
(\ref{eq:taueqnum2}) for the anisotropy torque, and 
choose $k_1$ and $k_2$ such that 
$(k_1 - k_2)/(k_1 + k_2) = 0.99$. This form of the anisotropy
is appropriate for thin magnetic layers, in which the magnetic
anisotropy is predominantly of easy-plane type. The magnitude
of the anisotropy energy is set by the parameters $q_{\rm f}$ and $j_{\rm f}$,
for which we take the same values as in the previous subsection.
All other parameters are also taken the same as in the previous
subsection.

The magnetization dynamics without applied magnetic field is
much richer than the magnetization dynamics at a large 
magnetic field. The reason is the existence of two stable 
equilibrium directions if no external magnetic field is applied
($\vm = \ve_3$ and $-\ve_3$). At sufficiently large current
densities, the current-induced torque drives the magnetization 
direction between these two stable directions, leading to a 
variety of dynamical phases.

\begin{figure}[t]
\epsfxsize=1\hsize
\hspace{0.05\hsize}
\epsffile{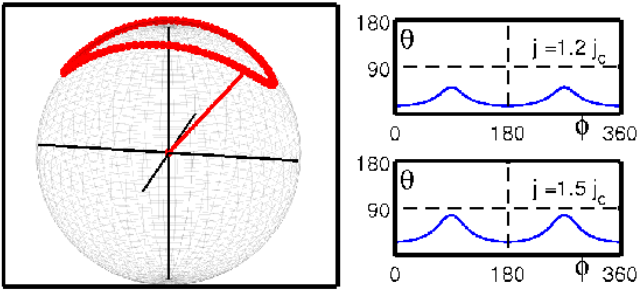}
\caption{\label{Fig:Disp1}
Typical elliptical trajectory for one of the
discrete nanomagnets $\mathbf{m}(n)$ for weak easy axis and strong
easy plane anisotropy with $j_{\rm c} < j < 2 j_{\rm c}$ (left panel). 
The upper and lower
right panels show the corresponding Poincar\'e sections for $j = 1.2 j_{\rm c}$
and $1.5 j_{\rm c}$ respectively.  This regime agrees with the perturbative
calculation of Sec.\ \protect\ref{sec:2}, where the lowest energy spin-wave
mode is excited and increasing the current only changes the amplitude
of elliptical oscillation.}
\end{figure}

For the numerical parameters chosen in our simulation, we observe
the following characteristic dynamical modes: For current densities
$j_{\rm c} < j \lesssim 2 j_{\rm c}$ the instability develops 
with the wavenumber 
$q=\pi/W_y$. Because the magnetic anisotropy energy used for the
simulation has no rotation symmetry around the $3$ axis, the
magnetization direction $\vm(n)$ of each magnet $n=1,\ldots,N$
traces out an ellipse, rather than a circle. We describe the
magnetization motion is described using Poincar\'e sections
for the polar angles $\theta$ and $\phi$ for the magnetization. 
The top right panel in Fig.~\ref{Fig:Disp1} shows traces that are
symmetric about $\phi = \pi$, which have the functional form for
$\mathbf{m}$ as predicted by the perturbation theory in the preceding
section.

\begin{figure}[t]
\epsfxsize=1\hsize
\hspace{0.05\hsize}
\epsffile{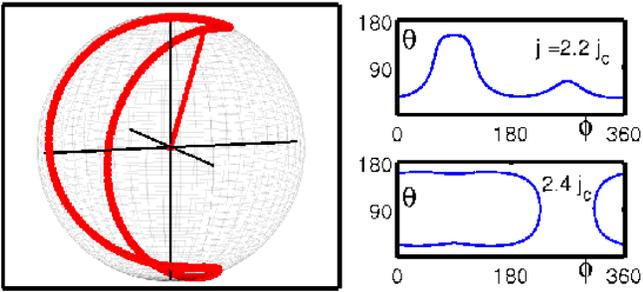}
\caption{\label{Fig:Disp2}
First manifestations of further dynamical instabilities in the
range $2 j_{\rm c} < j < 2.5 j_{\rm c}$. The upper right 
panel shows a Poincar\'e section for $j = 2.2 j_{\rm c}$ where the 
motion is no longer symmetric about the
easy axis.  The lower right panel shows the motion for $j = 2.4 j_{\rm c}$ where 
the motion is trapped between the $\pm \hat{e}_3$ easy axes direction.  
The left panel shows what this motion looks like on the unit sphere.}
\end{figure}

For higher currents with $2 j_{\rm c} \lesssim j \lesssim 2.5 j_{\rm c}$, the 
reflection symmetry about the easy axis is spontaneously broken,
resulting in asymmetric
ellipses (upper inset in Fig.\ \ref{Fig:Disp2}), 
which for even higher current densities turn into orbits
around the direction perpendicular to the easy axis (lower inset
in Fig.\ \ref{Fig:Disp2}). A three-dimensional rendering of this
regime is shown in Figure~\ref{Fig:Disp2}.

\begin{figure}[t]
\epsfxsize=1\hsize
\hspace{0.05\hsize}
\epsffile{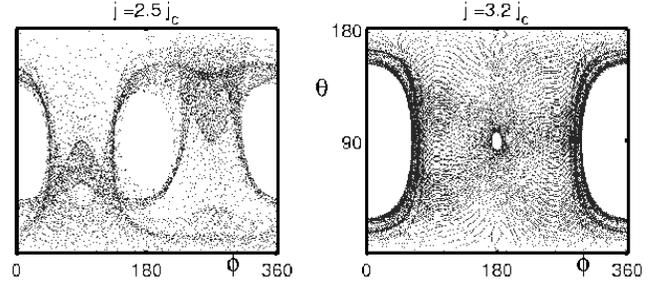}
\caption{\label{Fig:Disp3}
Poincar\'e sections for the magnetization direction of one of
the magnets at $j = 2.5 j_{\rm c}$ (left) and $j=3.2 j_{\rm c}$ (right).}
\end{figure}

For even larger currents there is a transition into non-periodic
modes that cover a significant part of phase space, as shown in 
Figure~\ref{Fig:Disp3}. Whereas these modes are non-ergodic for
lower current densities, they eventually become ergodic and
chaotic at high current densities, with 
Lyapunov exponents increasing with the current density $j$ (data
not shown).

\begin{figure}[t]
\epsfxsize=0.9\hsize
\hspace{0.05\hsize}
\epsffile{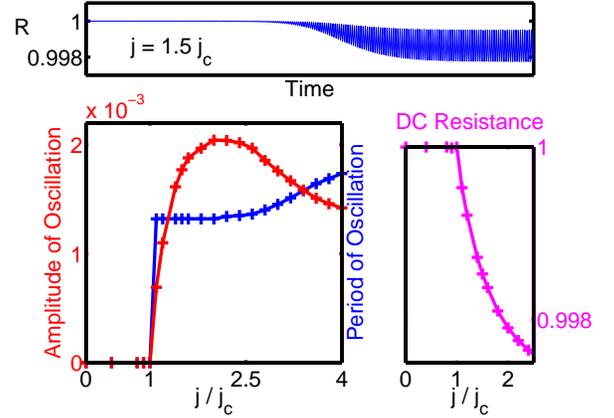}
\caption{\label{Fig:Disp4}
The upper panel shows the time trace of resistance where the spin-wave
instability causes a decrease in the observed resistance.  The lower left
plot shows how the amplitude and period of the resistance oscillation 
change with the driving current, while the lower right panel shows
the decrease of dc resistance.}
\end{figure}

In this general case, when the magnetization motion is not
just simple circular precession, the spin-wave instability
not only leads to a decrease of the dc resistance of the
ferromagnetic layer, it also causes a fast oscillation of
the resistance as shown in the time trace in Figure~\ref{Fig:Disp4}.
The right panel in Fig.\ \ref{Fig:Disp4} shows the decrease of
the dc resistance up to $j = 2.5 j_{\rm c}$. (No sufficiently
accurate numerical results were obtained for larger current
density.) Results for the variation of the resistance amplitude and
frequency with the applied current density are shown in the left 
panel for current densities up to $4 j_{\rm c}$. At the
parameter values considered in our simulation, the onset of 
the non-periodic magnetization variations is accompanied by a
sharp rise in precession frequency and a decrease of the 
amplitude of the resistance fluctuations.

\section{Conclusion}      

We have presented a detailed study of the transverse spin-wave
instability for a single ferromagnetic layer subject to a large
current perpendicular to the layer. Our calculations have been
in the small-amplitude regime, where perturbation theory can be
used, and in the large-amplitude regime, where the magnetization
dynamics can be solved numerically.

The two main signatures of the spin-wave instability are 
(1) existence of the instability for one current direction
only, and (2) a small
reduction in the dc resistance of the ferromagnetic layer. The
resistance
decrease arises because the existence of a spin wave with 
large amplitude lowers the spin accumulation in the normal 
metal adjacent to the ferromagnet. A lower spin accumulation
corresponds to a lower resistance (just as a high spin 
accumulation state of the antiparallel configuration in the 
standard current-perpendicular-to-plane giant magnetoresistance 
geometry gives a high resistance state). Both features have
been seen in a recent experiment.\cite{kn:oezyilmaz2004}

An important question for a dynamical instability is whether or
not it is hysteretic. Our calculation has shown that the
instability studied here is not, if a large magnetic field is
applied. Without applied magnetic field, the nature of the
spin wave instability depends on the precise form of the
magnetic anisotropy, and both hysteretic and non-hysteretic
behavior can be expected, in principle.

A noteworthy aspect of our calculation is that the spin-transfer
torque is calculated self-consistently: the magnitude and direction of
the spin-transfer torque depends on the spin accumulation in the
normal metal, which, in turn, depends on the precise magnetization
profile of the ferromagnet. In doing this, our work connects the
the circuit theory for hybrid ferromagnet--normal-metal
systems, which has been used extensively to describe 
the magnet's effect on spin 
accumulations in macrospin approximation,\cite{kn:tserkovnyak2005}
and micromagnetic
simulations, which, to date, have been restricted to simplified
models for the spin-transfer torque. However, our simulations should
be considered a proof-of-principle. They lack the spatial resolution 
and sophistication that full-scale micromagnetic simulations have.

\acknowledgments

We would like to thank A. Kent, I.\ Krivotorov, 
B.\ \"Ozyilmaz, D.\ Ralph, and
Ya.\ Tserkovnyak for discussions. This work was supported
by the Cornell Center for Materials research under NSF grant no.\
DMR 0079992, the Cornell Center for Nanoscale Systems under NSF 
grant no.\ EEC-0117770, by the NSF
under grant no.\ DMR 0334499, and by the Packard Foundation.

\appendix

\section*{Appendix}

The perturbative calculation of Sec.\ \ref{sec:2} focused on the case 
of a large applied magnetic field because, in that case, theoretical
results for the spin wave amplitude do not depend on sample-dependent
anisotropy energies. In this appendix we outline the theory for the
general case.

For the most general case, one needs a better ansatz for the
intrinsic torque $\vectau_{\rm an}$ than Eqs.\ (\ref{eq:tauint}) or
(\ref{eq:taumagneticfield}), as well as an expression for the 
current-induced
torque that does not rely on rotation symmetry around the easy 
axis. The general expression for the torque $\vectau_{\rm an}$
is most conveniently derived from the Free energy, $\vectau_{\rm an}
= -(\gamma/M) (\partial F/\partial \vm) \times \vm$. Since we are 
interested in the mode $\vm(\vq_{\rm c})$ only, we can expand $F$ in 
powers of $m_1(\vq_{\rm c})$ and $m_2(\vq_{\rm c})$, up to fourth oder as
\begin{eqnarray}
  F(\vm) &=& \frac{1}{2} \left[ k_1 m_1(\vq_{\rm c})^2 + k_2 m_2(\vq_{\rm c})^2
  \right] 
  \nonumber \\ && \mbox{}
  + \sum_{j=0}^{3} k_j' m_1(\vq_{\rm c})^j m_2(\vq_{\rm c})^{3-j}
  \nonumber \\ && \mbox{}
  + \sum_{j=0}^{4} k_j'' m_1(\vq_{\rm c})^j m_2(\vq_{\rm c})^{4-j}.
  \label{eq:Fexp}
\end{eqnarray}
The higher-order expansion constants $k_j'$, $j=0,1,2,3$, and
$k_j''$, $j=0,1,\ldots,4$, are not governed by any
special symmetry and therefore likely to be sample specific. 
(The cubic terms in the expansion of $F(\vm)$ may be forbidden
if there is a reflection symmetry around the easy axis. However, there
is no such symmetry in the presence of an applied magnetic field 
that is not aligned with the one of the sample's easy or hard axes,
so that cubic terms need to be included in a general treatment.)
Note that the higher-order torque terms are as important in
determining the
spin wave amplitude as the higher-order current-induced spin-transfer
torque. Unless these coefficients are known independently, a
calculation of the spin wave amplitude has no predictive value
--- that was the reason why the main text addressed the case of
a large applied magnetic field.

We now list our general results for the second
and third order potentials and third-order spin-transfer 
torque. The symbols used are defined in Sec.\ \ref{sec:2} of
the main text. The second-order charge potential expansion
coefficients for the normal-metal spacer are
\begin{widetext}
\begin{eqnarray}
a_{\rm c}^{(2)}(\vq) &=&   ej \sum_\vqp 
  \left\{
  \frac{[\Go + \Gq - 2 \Gqp]D(\vqp) + 2 [\Go - \Gqp][\Gq - \Gqp][g_1 + \Gqp]}
       {D(\vqp) g_{\rm m}(0) g_{\rm m}(\vq)}  \right. \nonumber \\
  && \ \ \ \ \mbox{} \times [m_1(\vqp) m_1(\vq -\vqp) + m_2(\vqp) 
  m_2(\vq - \vqp)] \nonumber \\ && \ \left. \mbox{} - 
   \frac{2 g_2 [\Go - \Gq][\Gq - \Gqp]}{D(\vq') g_{\rm m}(0) g_{\rm m}(\vq)} 
   [m_1(\vqp) m_2(\vq -\vqp) - m_2(\vqp) m_1(\vq - \vqp)] 
  \right\}
  \nonumber \\
&& - \hbar  \sum_{\vqp}
  \left\{ \frac{g_2 [\Gq - \Gqp] \Gq}{D(\vqp) g_{\rm m}(\vq)} 
  [m_1(\vq - \vqp) \dot{m}_1(\vqp) + m_2(\vq - \vqp) \dot{m}_2(\vqp)] 
  \right. \nonumber \\ && \ \left. \mbox{}
   +  \frac{[\Gq - \Gqp][D(\vqp) - \Gqp(g_1 + \Gqp)]}{D(\vqp) g_{\rm m}(\vq)} 
[m_1(\vq - \vqp) \dot{m}_2(\vqp) - m_2(\vq - \vqp) \dot{m}_1(\vqp)]
  \right\}. 
\end{eqnarray}
The coefficient $a_{\rm c}^{(2)}(0)$ determines how the
spin wave instability affects the resistance of the ferromagnetic
layer, cf.\ Eq.\ (\ref{eq:Rchange}) in the main text. The second order
correction to the $3$-component of the spin potential is given by
the expansion coefficients
\begin{eqnarray}
\va_{\rm s3}^{(2)}(\vq) &=& e j \sum_\vqp 
  \left\{ \frac{1}{2 g_{\rm m}(0)} 
  [m_1(\vqp) m_1(\vq - \vqp) + m_2(\vqp) m_2(\vq - \vqp)]
  \right. \nonumber \\ && \ \left. \mbox{}
  - \frac{1}{D(\vqp) g_{\rm m}(0) g_{\rm m}(\vq)}
  \left[D(\vqp) (g_+ + 2G_{\rm c}(\vq))(\Go + \Gq - 2\Gqp) 
  \right. \right. \nonumber \\ && \ \ \ \ \ \ \left. \left. \mbox{}
  - (\Go - \Gqp)(g_1 + \Gq) (g_{\rm m}(\vq) - 2 (g_+ + 2 G_{\rm c}(\vq)))
  (\Gq - \Gqp) \right] \right. \nonumber \\ && \ \ \ \ \mbox{} 
  \times [m_1(\vqp) m_1(\vq - \vqp) + m_2(\vqp) m_2(\vq - \vqp) ] 
  \nonumber \\  && \ \left. \mbox{}
  - \frac{g_2 [\Go - \Gq] [2(g_+ + 2G_{\rm c}(\vq))(\Gq - \Gqp) - g_{\rm m}(\vq)]}{D(\vqp) g_{\rm m}(0) g_{\rm m}(\vq)}
  [m_1(\vq) m_2(\vq -\vqp) - m_2(\vq) m_1(\vq - \vqp)] \right\}
  \nonumber \\ 
&& \mbox{} - \hbar \sum_{\vqp} \left\{
   \frac{g_2 \Gqp [g_{\rm m}(\vq) - 2(g_+ + 2
  G_{\rm c}(\vq))(\Gq - \Gqp)]}{2 D(\vqp) g_{\rm m}(\vq)}
   [m_1(\vq - \vqp) \dot{m}_1(\vqp) + m_2(\vq - \vqp) \dot{m}_2(\vqp)] 
  \right. \nonumber \\
&& \ \mbox{} - \frac{[D(\vqp) - \Gqp(g_1 + \Gqp)] [g_{\rm m}(\vq) - 2(g_+ + 2G_{\rm c}(\vq))(\Gq - \Gqp)]}{2 D(\vqp) g_{\rm m}(\vq)}
\nonumber \\
&& \left. \ \ \ \ \mbox{} [m_1(\vq - \vqp) \dot{m}_2(\vqp) - m_2(\vq - \vqp)
\dot{m}_1(\vqp) ] \vphantom{\frac{M}{M}}\right\}, 
\end{eqnarray}
The very first term describes the effect of a uniform magnetization
rotation; the remaining terms are the result of a non-uniform
magnetization. There are second-order corrections to the spin
potential expansion coefficients $a_1$ and $a_2$ that arise from
the presence of cubic terms in the anisotropy Free energy. Such
cubic terms cause second-order contributions to the time derivatives
$\dot{m}_1$ and $\dot{m}_2$, which give a contribution to the second
order spin potentials $a^{(2)}$ in the same way as 
the first-order time contribution to the time derivative affects
the first-order spin potentials $a^{(1)}$, see
Eq.\ (\ref{eq:afirst}).

Instead of listing the third-order potentials $a^{(3)}$, we describe
the corresponding current-induced torque. We specialize to the
contributions that are cubic in the magnetization amplitude at
wavevector $\vq_{\rm c}$. The resulting torque has terms
proportional to the third-order contributions to the time 
derivatives of the magnetization. These terms give rise to a
renormalized Gilbert damping coefficient and a renormalized
gyromagnetic ratio, see Eq.\ (\ref{eq:aren}). The remaining
terms can be written as $2\tilde \vectau(0) + \tilde \vectau(2
\vq_{\rm c})$, where (again using two-component spinor notation)
\begin{eqnarray}
  \tilde \tau^{(3)}(\vk) &=&
  \frac{- \hbar \gamma}{2 M d e^2 D(\vq_{\rm c})^2 g_{\rm m}(0) g_{\rm m}(\vk)}
  \nonumber \\ && \mbox{} \times \left\{
  2 e j g_2 D(\vq_{\rm c})
  \Gqc [\Gqc - \Gk][g_+ + 2 G_{\rm c}(\vk)] [\Go - \Gqc] 
  i \sigma_2 m(\vq_{\rm c}) m^{\rm T}(\vq_{\rm c}) m(\vq_{\rm c})
  \right. \nonumber \\
  && \ \mbox{} - 
  [g_{\rm m}(\vk) + 2(g_+ + 2G_{\rm c}(\vk))(\Gqc - \Gk)]
  \nonumber \\ && 
  \mbox{} \times \big[ -ej 
  [(2 \Gqc - \Go - \Gk) D(\vq_{\rm c}) 
  + 2(\Go - \Gqc)(g_1 + \Gqc)(\Gqc - \Gk)] 
  \nonumber \\ && \ \ \ \ \mbox{} \times  
  [ D(\vq_{\rm c}) - \Gqc (g_1 + \Gqc)] 
  m(\vq_{\rm c}) m^{\rm T}(\vq_{\rm c}) m(\vq_{\rm c})
\nonumber \\
&& \ \mbox{} + ej g_2 \Gqc [\Gqc - \Gk] [D(\vq_{\rm c}) + 
          2(\Go - \Gqc)(g_1 + \Gqc)] 
  i \sigma_2 m(\vq_{\rm c}) m^{\rm T}(\vq_{\rm c}) m(\vq_{\rm c})
\nonumber \\
&& \ \mbox{} + \hbar g_2 \Gqc g_{\rm m}(0) [\Gqc - \Gk] [D(\vq_{\rm c}) -
  \Gqc(g_1 + \Gqc)] \dot{m}(\vq_{\rm c}) m^{\rm T}(\vq_{\rm c}) m(\vq_{\rm c})
  \nonumber \\
&& \ \mbox{} + (\hbar/2) [\Gqc - \Gk] D(\vq_{\rm c}) g_{\rm m}(0) [D(\vq_{\rm
  c}) - \Gqc(2 g_1 + \Gqc)]
  \nonumber \\ && \ \ \ \ \mbox{} \times
  [ m(\vq_{\rm c}) m^{\rm T}(\vq_{\rm c}) i \sigma_2 \dot{m}(\vq_{\rm c})
  - i \sigma_2 m(\vq_{\rm c}) m^{\rm T}(\vq_{\rm c}) \dot{m}(\vq_{\rm c})]
  \nonumber \\ && \ \mbox{} 
  + (\hbar/2) [\Gqc - \Gk] g_{\rm m}(0) [(D(\vq_{\rm c}) - \Gqc(g_1 + \Gqc))^2 - (g_2 \Gqc)^2]
  i \sigma_2 \dot{m}(\vq_{\rm c}) m^{\rm T}(\vq_{\rm c}) m(\vq_{\rm c})
  \left. \big] \right\}. \nonumber \\
\end{eqnarray}
\end{widetext}

Once the perturbative expansions for the anisotropy torque and the
current-induced spin-transfer torque are known, the
Landau-Lifshitz-Gilbert equation can be solved for the magnetization
dynamics.


\begin{thebibliography}{41}
\expandafter\ifx\csname natexlab\endcsname\relax\def\natexlab#1{#1}\fi
\expandafter\ifx\csname bibnamefont\endcsname\relax
  \def\bibnamefont#1{#1}\fi
\expandafter\ifx\csname bibfnamefont\endcsname\relax
  \def\bibfnamefont#1{#1}\fi
\expandafter\ifx\csname citenamefont\endcsname\relax
  \def\citenamefont#1{#1}\fi
\expandafter\ifx\csname url\endcsname\relax
  \def\url#1{\texttt{#1}}\fi
\expandafter\ifx\csname urlprefix\endcsname\relax\def\urlprefix{URL }\fi
\providecommand{\bibinfo}[2]{#2}
\providecommand{\eprint}[2][]{\url{#2}}

\bibitem[{\citenamefont{Slonczewski}(1996)}]{kn:slonczewski1996}
\bibinfo{author}{\bibfnamefont{J.~C.} \bibnamefont{Slonczewski}},
  \bibinfo{journal}{J. Magn. Magn. Mater.} \textbf{\bibinfo{volume}{159}},
  \bibinfo{pages}{1} (\bibinfo{year}{1996}).

\bibitem[{\citenamefont{Berger}(1996)}]{kn:berger1996}
\bibinfo{author}{\bibfnamefont{L.}~\bibnamefont{Berger}},
  \bibinfo{journal}{Phys. Rev. B} \textbf{\bibinfo{volume}{54}},
  \bibinfo{pages}{9353} (\bibinfo{year}{1996}).

\bibitem[{\citenamefont{Tsoi et~al.}(1998)\citenamefont{Tsoi, Jansen, Bass,
  Chiang, Seck, Tsoi, and Wyder}}]{kn:tsoi1998}
\bibinfo{author}{\bibfnamefont{M.}~\bibnamefont{Tsoi}},
  \bibinfo{author}{\bibfnamefont{A.~G.~M.} \bibnamefont{Jansen}},
  \bibinfo{author}{\bibfnamefont{J.}~\bibnamefont{Bass}},
  \bibinfo{author}{\bibfnamefont{W.-C.} \bibnamefont{Chiang}},
  \bibinfo{author}{\bibfnamefont{M.}~\bibnamefont{Seck}},
  \bibinfo{author}{\bibfnamefont{V.}~\bibnamefont{Tsoi}}, \bibnamefont{and}
  \bibinfo{author}{\bibfnamefont{P.}~\bibnamefont{Wyder}},
  \bibinfo{journal}{Phys. Rev. Lett.} \textbf{\bibinfo{volume}{80}},
  \bibinfo{pages}{4281} (\bibinfo{year}{1998}).

\bibitem[{\citenamefont{Sun}(1999)}]{kn:sun1999}
\bibinfo{author}{\bibfnamefont{J.~Z.} \bibnamefont{Sun}}, \bibinfo{journal}{J.\
  Magn.\ Magn.\ Mater.} \textbf{\bibinfo{volume}{202}}, \bibinfo{pages}{157}
  (\bibinfo{year}{1999}).

\bibitem[{\citenamefont{Wegrowe et~al.}(1999)\citenamefont{Wegrowe, Kelly,
  Jaccard, Guittienne, and Ansermet}}]{kn:wegrowe1999}
\bibinfo{author}{\bibfnamefont{J.-E.} \bibnamefont{Wegrowe}},
  \bibinfo{author}{\bibfnamefont{D.}~\bibnamefont{Kelly}},
  \bibinfo{author}{\bibfnamefont{Y.}~\bibnamefont{Jaccard}},
  \bibinfo{author}{\bibfnamefont{P.}~\bibnamefont{Guittienne}},
  \bibnamefont{and} \bibinfo{author}{\bibfnamefont{J.-P.}
  \bibnamefont{Ansermet}}, \bibinfo{journal}{Europhys. Lett.}
  \textbf{\bibinfo{volume}{45}}, \bibinfo{pages}{626} (\bibinfo{year}{1999}).

\bibitem[{\citenamefont{Myers et~al.}(1999)\citenamefont{Myers, Ralph, Katine,
  Louie, and Buhrman}}]{kn:myers1999}
\bibinfo{author}{\bibfnamefont{E.}~\bibnamefont{Myers}},
  \bibinfo{author}{\bibfnamefont{D.}~\bibnamefont{Ralph}},
  \bibinfo{author}{\bibfnamefont{J.}~\bibnamefont{Katine}},
  \bibinfo{author}{\bibfnamefont{R.}~\bibnamefont{Louie}}, \bibnamefont{and}
  \bibinfo{author}{\bibfnamefont{R.}~\bibnamefont{Buhrman}},
  \bibinfo{journal}{Science} \textbf{\bibinfo{volume}{285}},
  \bibinfo{pages}{867} (\bibinfo{year}{1999}).

\bibitem[{\citenamefont{Katine et~al.}(2000)\citenamefont{Katine, Albert,
  Buhrman, Myers, and Ralph}}]{kn:katine2000}
\bibinfo{author}{\bibfnamefont{J.~A.} \bibnamefont{Katine}},
  \bibinfo{author}{\bibfnamefont{F.~J.} \bibnamefont{Albert}},
  \bibinfo{author}{\bibfnamefont{R.~A.} \bibnamefont{Buhrman}},
  \bibinfo{author}{\bibfnamefont{E.~B.} \bibnamefont{Myers}}, \bibnamefont{and}
  \bibinfo{author}{\bibfnamefont{D.~C.} \bibnamefont{Ralph}},
  \bibinfo{journal}{Phys. Rev. Lett.} \textbf{\bibinfo{volume}{84}},
  \bibinfo{pages}{3149} (\bibinfo{year}{2000}).

\bibitem[{\citenamefont{Kiselev et~al.}(2003)\citenamefont{Kiselev, Sankey,
  Krivorotov, Emley, Schoelkopf, and Buhrman}}]{kn:kiselev2003}
\bibinfo{author}{\bibfnamefont{S.~I.} \bibnamefont{Kiselev}},
  \bibinfo{author}{\bibfnamefont{J.~C.} \bibnamefont{Sankey}},
  \bibinfo{author}{\bibfnamefont{I.~N.} \bibnamefont{Krivorotov}},
  \bibinfo{author}{\bibfnamefont{N.~C.} \bibnamefont{Emley}},
  \bibinfo{author}{\bibfnamefont{R.~J.} \bibnamefont{Schoelkopf}},
  \bibnamefont{and} \bibinfo{author}{\bibfnamefont{R.~A.}
  \bibnamefont{Buhrman}}, \bibinfo{journal}{Nature}
  \textbf{\bibinfo{volume}{425}}, \bibinfo{pages}{380} (\bibinfo{year}{2003}).

\bibitem[{\citenamefont{Kiselev et~al.}(2004)\citenamefont{Kiselev, Sankey,
  Krivorotov, Emley, Rinkoski, Perez, Buhrman, and Ralph}}]{kn:kiselev2004}
\bibinfo{author}{\bibfnamefont{S.~I.} \bibnamefont{Kiselev}},
  \bibinfo{author}{\bibfnamefont{J.~C.} \bibnamefont{Sankey}},
  \bibinfo{author}{\bibfnamefont{I.~N.} \bibnamefont{Krivorotov}},
  \bibinfo{author}{\bibfnamefont{N.~C.} \bibnamefont{Emley}},
  \bibinfo{author}{\bibfnamefont{M.}~\bibnamefont{Rinkoski}},
  \bibinfo{author}{\bibfnamefont{C.}~\bibnamefont{Perez}},
  \bibinfo{author}{\bibfnamefont{R.~A.} \bibnamefont{Buhrman}},
  \bibnamefont{and} \bibinfo{author}{\bibfnamefont{D.~C.} \bibnamefont{Ralph}},
  \bibinfo{journal}{Phys. Rev. Lett.} \textbf{\bibinfo{volume}{93}},
  \bibinfo{eid}{036601} (\bibinfo{year}{2004}).

\bibitem[{\citenamefont{Rippard et~al.}(2004)\citenamefont{Rippard, Pufall,
  Kaka, Russek, and Silva}}]{kn:rippard2004}
\bibinfo{author}{\bibfnamefont{W.~H.} \bibnamefont{Rippard}},
  \bibinfo{author}{\bibfnamefont{M.~R.} \bibnamefont{Pufall}},
  \bibinfo{author}{\bibfnamefont{S.}~\bibnamefont{Kaka}},
  \bibinfo{author}{\bibfnamefont{S.~E.} \bibnamefont{Russek}},
  \bibnamefont{and} \bibinfo{author}{\bibfnamefont{T.~J.} \bibnamefont{Silva}},
  \bibinfo{journal}{Phys. Rev. Lett.} \textbf{\bibinfo{volume}{92}},
  \bibinfo{eid}{027201} (\bibinfo{year}{2004}).

\bibitem[{\citenamefont{Krivorotov et~al.}(2004)\citenamefont{Krivorotov,
  Emley, Garcia, Sankey, Kiselev, Ralph, and Buhrman}}]{kn:krivorotov2004}
\bibinfo{author}{\bibfnamefont{I.~N.} \bibnamefont{Krivorotov}},
  \bibinfo{author}{\bibfnamefont{N.~C.} \bibnamefont{Emley}},
  \bibinfo{author}{\bibfnamefont{A.~G.~F.} \bibnamefont{Garcia}},
  \bibinfo{author}{\bibfnamefont{J.~C.} \bibnamefont{Sankey}},
  \bibinfo{author}{\bibfnamefont{S.~I.} \bibnamefont{Kiselev}},
  \bibinfo{author}{\bibfnamefont{D.~C.} \bibnamefont{Ralph}}, \bibnamefont{and}
  \bibinfo{author}{\bibfnamefont{R.~A.} \bibnamefont{Buhrman}},
  \bibinfo{journal}{Phys. Rev. Lett.} \textbf{\bibinfo{volume}{93}},
  \bibinfo{eid}{166603} (\bibinfo{year}{2004}).

\bibitem[{\citenamefont{Waintal et~al.}(2000)\citenamefont{Waintal, Myers,
  Brouwer, and Ralph}}]{kn:waintal2000}
\bibinfo{author}{\bibfnamefont{X.}~\bibnamefont{Waintal}},
  \bibinfo{author}{\bibfnamefont{E.}~\bibnamefont{Myers}},
  \bibinfo{author}{\bibfnamefont{P.}~\bibnamefont{Brouwer}}, \bibnamefont{and}
  \bibinfo{author}{\bibfnamefont{D.}~\bibnamefont{Ralph}},
  \bibinfo{journal}{Phys. Rev. B} \textbf{\bibinfo{volume}{62}},
  \bibinfo{pages}{12317} (\bibinfo{year}{2000}).

\bibitem[{\citenamefont{Stiles and Zangwill}(2002)}]{kn:stiles2002}
\bibinfo{author}{\bibfnamefont{M.~D.} \bibnamefont{Stiles}} \bibnamefont{and}
  \bibinfo{author}{\bibfnamefont{A.}~\bibnamefont{Zangwill}},
  \bibinfo{journal}{Phys. Rev. B} \textbf{\bibinfo{volume}{66}},
  \bibinfo{pages}{014407} (\bibinfo{year}{2002}).

\bibitem[{\citenamefont{Xia et~al.}(2002)\citenamefont{Xia, Kelly, Bauer,
  Brataas, and Turek}}]{kn:xia2002}
\bibinfo{author}{\bibfnamefont{K.}~\bibnamefont{Xia}},
  \bibinfo{author}{\bibfnamefont{P.~J.} \bibnamefont{Kelly}},
  \bibinfo{author}{\bibfnamefont{G.~E.~W.} \bibnamefont{Bauer}},
  \bibinfo{author}{\bibfnamefont{A.}~\bibnamefont{Brataas}}, \bibnamefont{and}
  \bibinfo{author}{\bibfnamefont{I.}~\bibnamefont{Turek}},
  \bibinfo{journal}{Phys. Rev. B} \textbf{\bibinfo{volume}{65}},
  \bibinfo{eid}{220401} (\bibinfo{year}{2002}).

\bibitem[{\citenamefont{Tserkovnyak et~al.}(2004)\citenamefont{Tserkovnyak,
  Brataas, Bauer, and Halperin}}]{kn:tserkovnyak2005}
\bibinfo{author}{\bibfnamefont{Y.}~\bibnamefont{Tserkovnyak}},
  \bibinfo{author}{\bibfnamefont{A.}~\bibnamefont{Brataas}},
  \bibinfo{author}{\bibfnamefont{G.~E.~W.} \bibnamefont{Bauer}},
  \bibnamefont{and} \bibinfo{author}{\bibfnamefont{B.~I.}
  \bibnamefont{Halperin}}, \bibinfo{journal}{cond-mat/0409242}
  (\bibinfo{year}{2004}).

\bibitem[{\citenamefont{Sun}(2000)}]{kn:sun2000}
\bibinfo{author}{\bibfnamefont{J.~Z.} \bibnamefont{Sun}},
  \bibinfo{journal}{Phys. Rev. B} \textbf{\bibinfo{volume}{62}},
  \bibinfo{pages}{570} (\bibinfo{year}{2000}).

\bibitem[{\citenamefont{Bazaliy et~al.}(2004)\citenamefont{Bazaliy, Jones, and
  Zhang}}]{kn:bazaliy2004}
\bibinfo{author}{\bibfnamefont{Y.~B.} \bibnamefont{Bazaliy}},
  \bibinfo{author}{\bibfnamefont{B.~A.} \bibnamefont{Jones}}, \bibnamefont{and}
  \bibinfo{author}{\bibfnamefont{S.-C.} \bibnamefont{Zhang}},
  \bibinfo{journal}{Phys. Rev. B} \textbf{\bibinfo{volume}{69}},
  \bibinfo{eid}{094421} (\bibinfo{year}{2004}).

\bibitem[{\citenamefont{Li and Zhang}(2004)}]{kn:li2004}
\bibinfo{author}{\bibfnamefont{Z.}~\bibnamefont{Li}} \bibnamefont{and}
  \bibinfo{author}{\bibfnamefont{S.}~\bibnamefont{Zhang}},
  \bibinfo{journal}{Phys.\ Rev.\ B} \textbf{\bibinfo{volume}{69}},
  \bibinfo{eid}{134416} (\bibinfo{year}{2004}).

\bibitem[{\citenamefont{Xiao et~al.}(2005)\citenamefont{Xiao, Zangwill, and
  Stiles}}]{kn:xiao2005}
\bibinfo{author}{\bibfnamefont{J.}~\bibnamefont{Xiao}},
  \bibinfo{author}{\bibfnamefont{A.}~\bibnamefont{Zangwill}}, \bibnamefont{and}
  \bibinfo{author}{\bibfnamefont{M.~D.} \bibnamefont{Stiles}},
  \bibinfo{journal}{cond-mat/0504142}  (\bibinfo{year}{2005}).

\bibitem[{\citenamefont{Apalkov and Visscher}(2004)}]{kn:apalkov2004}
\bibinfo{author}{\bibfnamefont{D.}~\bibnamefont{Apalkov}} \bibnamefont{and}
  \bibinfo{author}{\bibfnamefont{P.}~\bibnamefont{Visscher}},
  \bibinfo{journal}{cond-mat/0405305}  (\bibinfo{year}{2004}).

\bibitem[{\citenamefont{Kovalev et~al.}(2005)\citenamefont{Kovalev, Bauer, and
  Brataas}}]{kn:kovalev2005}
\bibinfo{author}{\bibfnamefont{A.}~\bibnamefont{Kovalev}},
  \bibinfo{author}{\bibfnamefont{G.}~\bibnamefont{Bauer}}, \bibnamefont{and}
  \bibinfo{author}{\bibfnamefont{A.}~\bibnamefont{Brataas}},
  \bibinfo{journal}{cond-mat/0504705}  (\bibinfo{year}{2005}).

\bibitem[{\citenamefont{Li and Zhang}(2003)}]{kn:li2002}
\bibinfo{author}{\bibfnamefont{Z.}~\bibnamefont{Li}} \bibnamefont{and}
  \bibinfo{author}{\bibfnamefont{S.}~\bibnamefont{Zhang}},
  \bibinfo{journal}{Phys.\ Rev.\ B} \textbf{\bibinfo{volume}{68}},
  \bibinfo{eid}{024404} (\bibinfo{year}{2003}).

\bibitem[{\citenamefont{Berkov and Gorn}(2005{\natexlab{a}})}]{kn:berkov2005}
\bibinfo{author}{\bibfnamefont{D.}~\bibnamefont{Berkov}} \bibnamefont{and}
  \bibinfo{author}{\bibfnamefont{N.}~\bibnamefont{Gorn}},
  \bibinfo{journal}{Phys.\ Rev.\ B} \textbf{\bibinfo{volume}{71}},
  \bibinfo{eid}{052403} (\bibinfo{year}{2005}{\natexlab{a}}).

\bibitem[{\citenamefont{Torres et~al.}(2005)\citenamefont{Torres, Lopez-Diaz,
  Martinez, Carpentieri, and Finocchio}}]{kn:torres2005}
\bibinfo{author}{\bibfnamefont{L.}~\bibnamefont{Torres}},
  \bibinfo{author}{\bibfnamefont{L.}~\bibnamefont{Lopez-Diaz}},
  \bibinfo{author}{\bibfnamefont{E.}~\bibnamefont{Martinez}},
  \bibinfo{author}{\bibfnamefont{M.}~\bibnamefont{Carpentieri}},
  \bibnamefont{and}
  \bibinfo{author}{\bibfnamefont{G.}~\bibnamefont{Finocchio}},
  \bibinfo{journal}{J. Magn. Mag. Mater.} \textbf{\bibinfo{volume}{286}},
  \bibinfo{pages}{381} (\bibinfo{year}{2005}).

\bibitem[{\citenamefont{Berkov and Gorn}(2005{\natexlab{b}})}]{kn:berkov2005b}
\bibinfo{author}{\bibfnamefont{D.}~\bibnamefont{Berkov}} \bibnamefont{and}
  \bibinfo{author}{\bibfnamefont{N.}~\bibnamefont{Gorn}},
  \bibinfo{journal}{cond-mat/0503754}  (\bibinfo{year}{2005}{\natexlab{b}}).

\bibitem[{\citenamefont{Brataas et~al.}(2000)\citenamefont{Brataas, Nazarov,
  and Bauer}}]{kn:brataas2000b}
\bibinfo{author}{\bibfnamefont{A.}~\bibnamefont{Brataas}},
  \bibinfo{author}{\bibfnamefont{Y.~V.} \bibnamefont{Nazarov}},
  \bibnamefont{and} \bibinfo{author}{\bibfnamefont{G.~E.~W.}
  \bibnamefont{Bauer}}, \bibinfo{journal}{Phys. Rev. Lett.}
  \textbf{\bibinfo{volume}{84}}, \bibinfo{pages}{2481} (\bibinfo{year}{2000}).

\bibitem[{\citenamefont{Tserkovnyak et~al.}(2002)\citenamefont{Tserkovnyak,
  Brataas, and Bauer}}]{kn:tserkovnyak2002}
\bibinfo{author}{\bibfnamefont{Y.}~\bibnamefont{Tserkovnyak}},
  \bibinfo{author}{\bibfnamefont{A.}~\bibnamefont{Brataas}}, \bibnamefont{and}
  \bibinfo{author}{\bibfnamefont{G.~E.~W.} \bibnamefont{Bauer}},
  \bibinfo{journal}{Phys. Rev. Lett.} \textbf{\bibinfo{volume}{88}},
  \bibinfo{pages}{117601} (\bibinfo{year}{2002}).

\bibitem[{\citenamefont{Polianski and Brouwer}(2004)}]{kn:polianski2004}
\bibinfo{author}{\bibfnamefont{M.~L.} \bibnamefont{Polianski}}
  \bibnamefont{and} \bibinfo{author}{\bibfnamefont{P.~W.}
  \bibnamefont{Brouwer}}, \bibinfo{journal}{Phys. Rev. Lett.}
  \textbf{\bibinfo{volume}{92}}, \bibinfo{pages}{026602}
  (\bibinfo{year}{2004}).

\bibitem[{\citenamefont{{B. \"{O}zyilmaz, A. D. Kent, J. Z. Sun, M. J. Rooks,
  R. H. Koch}}(2004)}]{kn:oezyilmaz2004}
\bibinfo{author}{\bibnamefont{{B. \"{O}zyilmaz, A. D. Kent, J. Z. Sun, M. J.
  Rooks, R. H. Koch}}}, \bibinfo{journal}{Phys. Rev Lett.}
  \textbf{\bibinfo{volume}{93}}, \bibinfo{pages}{176604}
  (\bibinfo{year}{2004}).

\bibitem[{\citenamefont{Ji et~al.}(2003)\citenamefont{Ji, Chien, and
  Stiles}}]{kn:ji2003}
\bibinfo{author}{\bibfnamefont{Y.}~\bibnamefont{Ji}},
  \bibinfo{author}{\bibfnamefont{C.~L.} \bibnamefont{Chien}}, \bibnamefont{and}
  \bibinfo{author}{\bibfnamefont{M.~D.} \bibnamefont{Stiles}},
  \bibinfo{journal}{Phys.\ Rev.\ Lett.} \textbf{\bibinfo{volume}{90}},
  \bibinfo{eid}{106601} (\bibinfo{year}{2003}).

\bibitem[{\citenamefont{Stiles et~al.}(2004)\citenamefont{Stiles, Xiao, and
  Zangwill}}]{kn:stiles2004}
\bibinfo{author}{\bibfnamefont{M.~D.} \bibnamefont{Stiles}},
  \bibinfo{author}{\bibfnamefont{J.}~\bibnamefont{Xiao}}, \bibnamefont{and}
  \bibinfo{author}{\bibfnamefont{A.}~\bibnamefont{Zangwill}},
  \bibinfo{journal}{Phys. Rev. B.} \textbf{\bibinfo{volume}{69}},
  \bibinfo{pages}{054408} (\bibinfo{year}{2004}).

\bibitem[{\citenamefont{\"Ozyilmaz et~al.}(2005)\citenamefont{\"Ozyilmaz, Kent,
  Rooks, and Sun}}]{kn:oezyilmaz2005}
\bibinfo{author}{\bibfnamefont{B.}~\bibnamefont{\"Ozyilmaz}},
  \bibinfo{author}{\bibfnamefont{A.~D.} \bibnamefont{Kent}},
  \bibinfo{author}{\bibfnamefont{M.~J.} \bibnamefont{Rooks}}, \bibnamefont{and}
  \bibinfo{author}{\bibfnamefont{J.~Z.} \bibnamefont{Sun}},
  \bibinfo{journal}{Phys.\ Rev.\ B} \textbf{\bibinfo{volume}{71}},
  \bibinfo{eid}{140403} (\bibinfo{year}{2005}).

\bibitem[{\citenamefont{Urazhdin}(2004)}]{kn:urazhdin2004}
\bibinfo{author}{\bibfnamefont{S.}~\bibnamefont{Urazhdin}},
  \bibinfo{journal}{Phys. Rev. B.} \textbf{\bibinfo{volume}{69}},
  \bibinfo{pages}{134430} (\bibinfo{year}{2004}).

\bibitem[{\citenamefont{Brataas et~al.}(2005)\citenamefont{Brataas,
  Tserkovnyak, and Bauer}}]{kn:brataas2005}
\bibinfo{author}{\bibfnamefont{A.}~\bibnamefont{Brataas}},
  \bibinfo{author}{\bibfnamefont{Y.}~\bibnamefont{Tserkovnyak}},
  \bibnamefont{and} \bibinfo{author}{\bibfnamefont{G.}~\bibnamefont{Bauer}},
  \bibinfo{journal}{cond-mat/0501672}  (\bibinfo{year}{2005}).

\bibitem[{\citenamefont{Slavin and Kabos}(2005)}]{kn:slavin2005}
\bibinfo{author}{\bibfnamefont{A.}~\bibnamefont{Slavin}} \bibnamefont{and}
  \bibinfo{author}{\bibfnamefont{P.}~\bibnamefont{Kabos}},
  \bibinfo{journal}{IEEE Trans. Mag.} \textbf{\bibinfo{volume}{41}},
  \bibinfo{pages}{1264} (\bibinfo{year}{2005}).

\bibitem[{\citenamefont{Chen et~al.}(2004)\citenamefont{Chen, Ji, Chien, and
  Stiles}}]{kn:chen2004}
\bibinfo{author}{\bibfnamefont{T.~Y.} \bibnamefont{Chen}},
  \bibinfo{author}{\bibfnamefont{Y.}~\bibnamefont{Ji}},
  \bibinfo{author}{\bibfnamefont{C.~L.} \bibnamefont{Chien}}, \bibnamefont{and}
  \bibinfo{author}{\bibfnamefont{M.~D.} \bibnamefont{Stiles}},
  \bibinfo{journal}{Phys.\ Rev.\ Lett.} \textbf{\bibinfo{volume}{93}},
  \bibinfo{eid}{026601} (\bibinfo{year}{2004}).

\bibitem[{\citenamefont{Stiles}(1996)}]{kn:stiles1996}
\bibinfo{author}{\bibfnamefont{M.~D.} \bibnamefont{Stiles}},
  \bibinfo{journal}{J. Appl. Phys.} \textbf{\bibinfo{volume}{79}},
  \bibinfo{pages}{5805} (\bibinfo{year}{1996}).

\bibitem[{\citenamefont{Lifschitz and Pitaevskii}(1980)}]{kn:lifshitz1980}
\bibinfo{author}{\bibfnamefont{E.~M.} \bibnamefont{Lifschitz}}
  \bibnamefont{and} \bibinfo{author}{\bibfnamefont{L.~P.}
  \bibnamefont{Pitaevskii}}, \emph{\bibinfo{title}{Statistical Physics, part
  2}} (\bibinfo{publisher}{Pergamon}, \bibinfo{year}{1980}).

\bibitem[{\citenamefont{Gilbert}(2004)}]{kn:gilbert2004}
\bibinfo{author}{\bibfnamefont{T.~L.} \bibnamefont{Gilbert}},
  \bibinfo{journal}{IEEE Trans.\ Mag.} \textbf{\bibinfo{volume}{40}},
  \bibinfo{pages}{3443} (\bibinfo{year}{2004}).

\bibitem[{\citenamefont{\mbox{Ya.} B.~Bazaliy
  et~al.}(1998)\citenamefont{\mbox{Ya.} B.~Bazaliy, Jones, and
  Zhang}}]{kn:bazaliy1998}
\bibinfo{author}{\bibnamefont{\mbox{Ya.} B.~Bazaliy}},
  \bibinfo{author}{\bibfnamefont{B.~A.} \bibnamefont{Jones}}, \bibnamefont{and}
  \bibinfo{author}{\bibfnamefont{S.-C.} \bibnamefont{Zhang}},
  \bibinfo{journal}{Phys. Rev. B} \textbf{\bibinfo{volume}{57}},
  \bibinfo{pages}{3213} (\bibinfo{year}{1998}).

\bibitem[{\citenamefont{Wohlfahrt}(1980)}]{kn:wohlfahrt1980}
\bibinfo{author}{\bibfnamefont{E.~P.} \bibnamefont{Wohlfahrt}}, in
  \emph{\bibinfo{booktitle}{Ferromagnetic Materials}}, edited by
  \bibinfo{editor}{\bibfnamefont{E.~P.} \bibnamefont{Wohlfahrt}}
  (\bibinfo{publisher}{North-Holland}, \bibinfo{year}{1980}),
  vol.~\bibinfo{volume}{1}.

\end{thebibliography}
\end{document}